\documentclass[aps,prd,preprint,nofootinbib,11pt]{revtex4}
\pdfoutput=1
\usepackage{amsfonts}
\usepackage{mathrsfs}
\usepackage{graphicx}
\usepackage{amsmath}
\usepackage{amssymb}
\usepackage{bm}
\usepackage{booktabs}
\usepackage{multirow}
\usepackage{ulem}
\usepackage{placeins}
\usepackage{float}
\usepackage{color}
\usepackage{slashed}
\usepackage{hyperref}
\allowdisplaybreaks[1]
\graphicspath{{figures/}}

\hypersetup{
  colorlinks=true,
  linkcolor=blue,
  citecolor=blue,
  urlcolor=blue
}

\newcommand{\MeV}{\mathrm{MeV}}
\newcommand{\GeV}{\mathrm{GeV}}

\begin{document}

\title{\Large Nodal filtering in open-charm decays of the $\psi(4040)$--$\psi(4160)$ system\\[7mm]}

\author{Bing-Dong Wan}
\email{wanbd@lnnu.edu.cn}
\affiliation{School of Physics and Electronic Technology, Liaoning Normal University, Dalian 116029, China}
\affiliation{Center for Theoretical and Experimental High Energy Physics, Liaoning Normal University, Dalian 116029, China}

\begin{abstract}
Hadronic decay channels can provide direct information on the momentum-space structure of confined quark systems. We investigate this possibility in the $\psi(4040)$--$\psi(4160)$ system using an instantaneous Bethe--Salpeter (BS) framework combined with a relativistic $^3P_0$ decay model. The BS solutions yield nearby bare $3\,^3S_1$ and $2\,^3D_1$ states at 4051 and 4110 MeV, respectively, whose opposite displacements from the physical vector states motivate an effective two-state level-repulsion description. By comparing the matched open-charm channels $D\bar D$, $D\bar D^*+{\rm c.c.}$, $D_s\bar D_s$, and $D_s\bar D_s^*+{\rm c.c.}$ with the same leading $P$-wave threshold behavior, we show that channel-dependent overlap kernels act as momentum-space nodal filters. At $M=4146$ MeV, the $PV/PP$ width ratio is 0.253 in the nonstrange sector but 1.63 in the strange sector, reversing the ordering expected from phase space alone. The momentum-resolved amplitudes reveal that different channels weight opposite sides of the same $2D$ nodal region differently, leading to distinct cancellation patterns and a common recoil condition, $P_f\simeq0.775$ GeV, for charge-resolved $D\bar D^*$ amplitude zeros. These results establish open-charm decays as a probe of momentum-space wave-function structures in heavy quarkonium.
\end{abstract}

\maketitle

\section{Introduction}

Exclusive hadronic decays provide not only measurements of strong-interaction rates but also information on the internal dynamics of excited hadrons. In particular, for radially and orbitally excited quarkonium states, nodes in the momentum-space wave functions can induce strong cancellations in decay amplitudes and lead to anomalous channel suppressions within quark-pair-creation approaches~\cite{Barnes2005,GodfreyIsgur}. However, previous studies have mainly focused on the impact of nodal structures on individual decay channels or overall width patterns. Whether different final-state channels can selectively probe different momentum regions of the same initial-state wave function and reveal correlated signatures of its internal structure remains an open question.

The established $\psi(4040)$ and $\psi(4160)$ states are conventionally assigned predominantly to the $3\,{}^3S_1$ and $2\,{}^3D_1$ charmonium configurations, respectively~\cite{Barnes2005,GodfreyIsgur,Brambilla}. Their nearby masses and rich open-charm decay patterns make them particularly sensitive to the relation between quarkonium wave functions and observable decay amplitudes. Relativistic dynamics and $S$--$D$ mixing can affect both their spectroscopic interpretation and their open-charm decay patterns~\cite{Ke2026,Man2025vmm}. Measurements of $e^+e^-\to D\bar D$, $D\bar D^*+\mathrm{c.c.}$, and $D_s^{(*)}\bar D_s^{(*)}$ in this energy region therefore provide information complementary to the mass spectrum~\cite{BESIII_DD_2024,BESIII_DstarD_2022,BESIII_DsDs_2024}. However, the possibility that different open-charm channels can act as distinct probes of the same momentum-space quarkonium wave function remains largely unexplored.

In this work, we investigate how the nearby $3S$ and $2D$ charmonium solutions obtained within a common relativistic BS framework manifest themselves in both the $\psi(4040)$--$\psi(4160)$ spectrum and channel-resolved open-charm decay amplitudes. The BS equation yields both the bare masses and the full multicomponent Salpeter wave functions. The bare $3S$ level lies above the $\psi(4040)$ region, whereas the bare $2D$ level lies below the $\psi(4160)$ region. This opposite displacement motivates an effective two-state description in terms of level repulsion. More importantly, the $2D$ wave function exhibits a nontrivial nodal structure with sign changes in momentum space. Since a strong-decay amplitude is a channel-dependent overlap integral, the same initial wave function can be filtered differently by overlap kernels with different spin, flavor, and recoil structures. This provides a direct connection between the internal momentum-space structure and measurable open-charm decay patterns.

We consider the four channels
\[
\{D\bar D,\;D\bar D^*+\mathrm{c.c.},\;D_s\bar D_s,\;D_s\bar D_s^*+\mathrm{c.c.}\}
\]
as a matched $2\times2$ flavor--spin channel set. We refer to them as $PP$ and $PV$ modes, with $P$ and $V$ denoting pseudoscalar and vector open-charm mesons. The $PP$--$PV$ comparison changes the final-state spin structure, while the nonstrange--strange comparison changes the light flavor; in both cases the corresponding recoil and final-state wave functions also change. All four modes have the lowest allowed relative orbital momentum $L=1$ and therefore share the same leading $|\bm P_f|^3$ threshold behavior. This common $P$-wave scaling removes differences associated with the leading threshold power, allowing the remaining channel dependence to probe the interplay of spin, flavor, recoil, and dynamical wave-function overlap.

Within this framework, the pure-$2D$ component is first analyzed through channel-resolved overlap diagnostics to trace the decay pattern back to the momentum-space structure of the initial wave function. The $3S$ and $2D$ amplitudes are subsequently combined coherently to examine the resulting interference pattern. We use $4146~\MeV$, motivated by the recent reanalysis $M_{\psi(4160)}=4145.76\pm4.48~\MeV$~\cite{Peng2024}, as the central benchmark.

The remainder of this paper is organized as follows. Following this Introduction, Section~\ref{sec:spectrum} presents the BS spectrum and the effective two-state description. Section~\ref{sec:formalism} introduces the common decay formalism and the matched channel basis. Sections~\ref{sec:results} and~\ref{sec:nodal} establish the pure-$2D$ channel structure and its momentum-space origin, while Sec.~\ref{sec:mixing_results} presents the coherent $3S$--$2D$ decay pattern. Section~\ref{sec:discussion} discusses how these results can be confronted with an amplitude analysis of the $\psi(4040)$--$\psi(4160)$ system. Detailed numerical diagnostics are collected in the appendices.

\section{Bethe--Salpeter spectrum and effective two-state description}
\label{sec:spectrum}

The charmonium and open-charm wave functions are obtained from the instantaneous Bethe--Salpeter equation with a screened Cornell-type kernel~\cite{Salpeter1952,SalpeterOrigins,Chang2005,Chang2004,Wang2006}. The equation, its Salpeter reduction, the interaction kernel, the energy projectors, and the complete $1^{--}$ wave-function decomposition are summarized in Appendix~\ref{app:bs}. The same construction determines both the spectrum and the wave functions entering the strong-decay amplitudes. For the vector states, the Salpeter solution contains four independent radial functions, which we denote by $f_3$, $f_4$, $f_5$, and $f_6$. The constrained relativistic components reconstructed from these functions enter the decay overlap on the same footing.

For the vector levels relevant to the $\psi(4040)$--$\psi(4160)$ system, we obtain
\begin{equation}
M_{3\,{}^3S_1}^{(0)}=4051~\MeV,
\qquad
M_{2\,{}^3D_1}^{(0)}=4110~\MeV.
\label{eq:bs_masses}
\end{equation}
The calculated $4S$ level lies at $4330~\MeV$ and is therefore well separated from the nearby $3S$--$2D$ pair considered here. Throughout this work, the superscript $(0)$ denotes the unmixed BS eigenvalues before the effective two-state construction introduced below. The bare $3S$ level lies above the $\psi(4040)$ region, whereas the bare $2D$ level lies below the $\psi(4160)$ region. This opposite displacement motivates, but does not by itself derive, an effective level-repulsion description of the two-state system.

To characterize this pattern phenomenologically, we introduce the minimal real symmetric mass matrix
\begin{equation}
\mathcal H_{SD}
=
\begin{pmatrix}
M_{3S}^{(0)}+\delta & V_{SD}\\
V_{SD} & M_{2D}^{(0)}+\delta
\end{pmatrix},
\label{eq:effective_mass_matrix}
\end{equation}
where $\delta$ is a common shift of the two-state centroid and $V_{SD}$ is an effective real off-diagonal mixing element. Taking
\begin{equation}
(M_L,M_H)=(4040,4146)~\MeV
\end{equation}
as a representative pair of physical masses gives
\begin{equation}
\delta=12.5~\MeV,
\qquad
|V_{SD}|=44.0~\MeV,
\qquad
|\theta|=28.1^\circ,
\label{eq:mass_derived_mixing}
\end{equation}
with
\begin{equation}
\tan 2\theta
=
\frac{2V_{SD}}
{M_{2D}^{(0)}-M_{3S}^{(0)}}.
\end{equation}
The common offset is fixed by the difference between the bare and physical centroids,
\begin{equation}
\delta
=
\frac{
M_L+M_H-M_{3S}^{(0)}-M_{2D}^{(0)}
}{2},
\end{equation}
while $V_{SD}$ accounts for the additional level splitting within this effective two-state parametrization.

Equation~\eqref{eq:effective_mass_matrix} is used only as a phenomenological organization of the BS spectrum. Neither $\delta$ nor $V_{SD}$ is calculated from the BS kernel; both are inferred by requiring the eigenvalues of Eq.~\eqref{eq:effective_mass_matrix} to reproduce the chosen physical masses. The resulting $|\theta|$ is therefore a mass-organized estimate rather than a dynamical prediction of the $S$--$D$ mixing angle. In particular, the mass spectrum determines only the magnitude of the mixing within this ansatz. In the fixed real phase convention adopted for the BS wave functions and decay amplitudes, it does not determine the relative sign entering their coherent combination.

We therefore parameterize the orthogonal lower and upper states as
\begin{equation}
|\psi_L\rangle
=
\cos\theta\,|3S\rangle
-\eta\sin\theta\,|2D\rangle,
\qquad
|\psi_H\rangle
=
\cos\theta\,|2D\rangle
+\eta\sin\theta\,|3S\rangle,
\label{eq:psi4160_admixture}
\end{equation}
where $\eta=\pm1$ labels the two relative-sign choices that remain unresolved by the mass spectrum within the fixed real convention adopted here. Their consequences are examined later at the amplitude level. Before introducing the coherent $3S$ contribution, we first analyze the pure-$2D$ component separately in order to isolate its momentum-space structure and the resulting channel-dependent decay pattern.

\section{Relativistic open-charm amplitudes and matched channels}
\label{sec:formalism}

The open-charm decay amplitudes are evaluated within a relativistic
$^3P_0$ model~\cite{Micu,LeYaouanc,Segovia2012,Wang2013Strong}
using the positive-energy Salpeter wave functions obtained in
Sec.~\ref{sec:spectrum}. Keeping the leading positive-energy contribution
in the instantaneous Mandelstam reduction, following the treatment in
Refs.~\cite{Wang2013Strong,Wan2026}, the transition amplitude for
$A(P)\rightarrow B(P_1)+C(P_2)$ is written as
\begin{equation}
\mathcal M_{a\lambda}^{(q)}
=
g_q
\int_0^\infty dq\, I_{a\lambda}(q),
\label{eq:amplitude_general}
\end{equation}
where $a$ denotes the charge sector and $\lambda$ labels the final-state
spin projection. The radial integrand is defined by
\begin{equation}
I_{a\lambda}(q)
=
\frac{q^2}{(2\pi)^3}
\int d\Omega_q\,
\mathrm{Tr}
\left[
\bar\varphi_{B,a}^{++}(\bm q_B)
\varphi_A^{++}(\bm q)
\bar\varphi_{C,a}^{++}(\bm q_C)
\right]_{\lambda}.
\label{eq:radial_integrand_definition}
\end{equation}
The recoil-shifted momenta and the complete Salpeter wave-function
decomposition entering the overlap are given in Appendix~\ref{app:bs}.
The numerical calculation retains the full recoil dependence and all
positive-energy Salpeter components. The charge assignments, polarization
conventions, and multiplicity factors used in the numerical evaluation
are summarized in Appendix~\ref{app:decay_conventions}. In particular,
Eq.~\eqref{eq:radial_integrand_definition} defines the radial integrand
used in the diagnostics below and is not factorized into an initial-state
radial wave function and a channel-dependent kernel.

The corresponding two-body partial width is
\begin{equation}
\Gamma_i
=
\frac{|\bm P_f|}{8\pi M_A^2}
\frac{1}{2J_A+1}
\sum_{\mathrm{pol.}}
|\mathcal M_i|^2 ,
\label{eq:width}
\end{equation}
where the polarization sum is performed after constructing the coherent
amplitudes. The overall pair-creation coefficient cancels in the
same-flavor ratios considered below, so that their channel dependence is
controlled by the overlap structure of the decay amplitudes.

To isolate the dynamical channel dependence, we consider the matched
open-charm channels
\begin{equation}
\begin{array}{c|cc}
&PP&PV\\
\hline
\mathrm{nonstrange}
&D\bar D&D\bar D^*+\mathrm{c.c.}\\
\mathrm{strange}
&D_s\bar D_s&D_s\bar D_s^*+\mathrm{c.c.}
\end{array}
\label{eq:channel_basis}
\end{equation}
where $P$ and $V$ denote pseudoscalar and vector open-charm mesons. The
$PP$--$PV$ comparison changes the final-state spin structure, while the
nonstrange--strange comparison changes the light flavor. The associated
recoil momenta and final-state wave functions are fully retained in each
channel.

All four channels have the lowest allowed relative orbital momentum
$L=1$ and therefore share the same leading threshold behavior,
\begin{equation}
\Gamma_i\propto |\bm P_f|^3 .
\end{equation}
This common $P$-wave scaling removes differences associated with the
leading threshold power and allows the remaining channel dependence to
probe the interplay of spin, flavor, recoil, and dynamical wave-function
overlap.

At fixed light flavor, we define the relative $PV$ responses
\begin{equation}
R_n=
\frac{\Gamma(D\bar D^*+\mathrm{c.c.})}
{\Gamma(D\bar D)},
\qquad
R_s=
\frac{\Gamma(D_s\bar D_s^*+\mathrm{c.c.})}
{\Gamma(D_s\bar D_s)} .
\label{eq:ratios}
\end{equation}
To characterize the relative flavor dependence of the
$PP\rightarrow PV$ response, we introduce the bounded asymmetry
\begin{equation}
A_{sn}
=
\frac{R_s-R_n}{R_s+R_n}.
\label{eq:bounded_asymmetry}
\end{equation}
The quantities $R_n$, $R_s$, and $A_{sn}$ are independent of the overall
pair-creation normalization. Additional cross-flavor ratios and
threshold-normalized quantities are collected in Appendix~\ref{app:inputs}.

For the mixed states introduced in Sec.~\ref{sec:spectrum}, the $3S$ and
$2D$ contributions are combined coherently at the amplitude level. For
each charge and polarization component, we use
\begin{equation}
\mathcal M_{a\lambda}(\theta)
=
\cos\theta\,
\mathcal M_{a\lambda}(2D)
+
\sin\theta\,
\mathcal M_{a\lambda}(3S),
\label{eq:coherent_amplitude}
\end{equation}
where the sign of $\theta$ specifies the relative sign between the two
components in the fixed real phase convention adopted for the basis
states. The polarization and charge sums are performed only after this
coherent amplitude combination.

\section{Numerical inputs}
\label{sec:numerical}

The model parameters follow our previous Bethe--Salpeter analyses of vector
charmonium decays~\cite{Wan2026,Chang2005,WangGL2020,Geng2019,Chang2015,DingWan}.
The charmonium and heavy-light wave functions are obtained within a common
screened Cornell setup. The complete parameter set and the corresponding
$V_0$ values are collected in Appendix~\ref{app:inputs}.

The nonstrange pair-creation vertex is calibrated from
$\psi(3770)\rightarrow D\bar D$. With
$m_n=(m_u+m_d)/2=0.308~\GeV$ and
$\gamma_n=0.526\pm0.023$, the coefficient entering the decay amplitude is
\begin{equation}
g_n=2m_n\gamma_n=0.324\pm0.014~\GeV .
\end{equation}

This corresponds to a $4.3\%$ uncertainty in the overall pair-creation
normalization. Since absolute widths scale as $g_n^2$, they inherit an
$8.6\%$ normalization uncertainty. The observables
$R_n$, $R_s$, and $A_{sn}$ are independent of this overall normalization.
The baseline strange prescription is taken as $g_s=g_n$; alternative
strange-pair normalizations affect absolute cross-flavor ratios but not
the same-flavor observables discussed here. Model-dependent uncertainties
from the kernel parameters and the positive-energy projection are not
included in the present uncertainty estimate.

We consider two external masses,
\begin{equation}
M_A\in\{4146,\,4191\}~\MeV .
\end{equation}
The $4146~\MeV$ point is taken as the central reference mass, while the
$4191~\MeV$ point provides a higher-recoil cross-check. 
The second point, $M_A=4191$ MeV, is not associated with an additional experimental mass determination. It is selected as a representative higher-recoil point within the $\psi(4160)$ region to test whether the nodal-filtering pattern survives under modified external kinematics.
We further perform
a continuous scan over
$M_A=4140$--$4200~\MeV$
to examine whether the channel ordering persists away from the central
point.

Throughout this scan, the tabulated radial wave functions are kept fixed.
Only the external kinematics, including phase space, recoil, and the
mass-dependent positive-energy coefficients, are updated. Therefore, the
scan represents a kinematic continuation of the decay amplitudes rather
than a sequence of self-consistent BS solutions.

\section{Pure-$2D$ channel structure}
\label{sec:results}

We first analyze the pure-$2D$ component at $4146~\MeV$, with $4191~\MeV$ retained as a higher-recoil cross-check. This isolates the nodal filtering, recoil dependence, and flavor--spin pattern generated by the $2D$ wave function across the matched open-charm channels. We then compare the matched-channel ratios with the corresponding $P$-wave threshold reference and trace the nonstrange $PV$ minimum to the charge-resolved amplitudes. Section~\ref{sec:mixing_results} subsequently combines the resulting $2D$ amplitudes coherently with the corresponding $3S$ amplitudes to obtain the decay pattern of the upper state.

\subsection{Pure-$2D$ decay pattern at two external masses}

The pure-$2D$ decay widths obtained with fixed radial wave functions are
summarized in Table~\ref{tab:widths_pure}. The $4146~\MeV$ point is taken
as the central reference mass, while $4191~\MeV$ is included only as a
higher-recoil cross-check. The radial wave functions are kept fixed, while
the external mass changes the phase space, recoil dependence, and the
positive-energy coefficient reconstruction.

\begin{table}[t]
\caption{Charge-summed decay widths of the pure-$2,{}^3D_1$ state at two
external masses in the matched $PP/PV$ basis. The radial wave functions are
fixed, while the external mass changes the decay kinematics and recoil
dependence.}
\label{tab:widths_pure}
\begin{ruledtabular}
\begin{tabular}{lcccc}
Channel & $|\bm P_f|$ at $4146$ & $\Gamma_i(4146)$ & $|\bm P_f|$ at $4191$ & $\Gamma_i(4191)$ \\
 & (MeV) & (MeV) & (MeV) & (MeV) \\
\hline
$D\bar D$ & 894.7 & 5.97 & 945.6 & 9.83 \\
$D\bar D^*+\mathrm{c.c.}$ & 730.2 & 1.51 & 791.7 & 0.44 \\
$D_s\bar D_s$ & 651.4 & 4.20 & 719.8 & 2.04 \\
$D_s\bar D_s^*+\mathrm{c.c.}$ & 368.2 & 6.85 & 478.8 & 7.43 \\
\end{tabular}
\end{ruledtabular}
\end{table}

At $M_A=4146~\MeV$, the two $PP$ widths are comparable, while the
nonstrange $D\bar D^*+\mathrm{c.c.}$ channel is suppressed and the strange
$D_s\bar D_s^*+\mathrm{c.c.}$ channel gives the largest width among the four
matched modes. At $M_A=4191~\MeV$, the nonstrange $PV$ channel becomes
strongly suppressed, whereas the strange $PV$ channel remains sizable. This
opposite behavior already indicates that the channel pattern is not
controlled by phase space alone.

\subsection{Ratios and threshold comparison}

We first test whether the calculated ratios follow the corresponding
charge-summed $|\bm P_f|^3$ threshold reference. Table~\ref{tab:ratios_phase_space}
compares the full decay calculation with this purely kinematic expectation.
At the central point $M_A=4146~\MeV$, the full result gives
\begin{equation}
(R_n,R_s,A_{sn})=(0.2529,1.6298,0.731),
\end{equation}
whereas the threshold reference gives
$A_{sn}=-0.503$, corresponding to the opposite ordering
$R_s<R_n$. The full-calculation ordering remains unchanged across the
fixed-wave-function scan and is also present at the $4191~\MeV$
higher-recoil cross-check. The reversal relative to the threshold
reference therefore indicates dynamical overlap effects beyond the common
leading threshold factor.

Additional cross-flavor ratios, reduced strengths, and strange-pair
normalization tests are collected in Appendix~\ref{app:inputs}.

\begin{table}[t]
\caption{
Same-flavor spin-replacement ratios $R_n$, $R_s$, and the bounded
asymmetry $A_{sn}$ at two external masses. The full decay calculation is
compared with the charge-summed $|\bm P_f|^3$ threshold reference, which
contains only the leading $P$-wave phase-space scaling.
}
\label{tab:ratios_phase_space}
\begin{ruledtabular}
\begin{tabular}{lcccc}
&
\multicolumn{2}{c}{$M_A=4146~\MeV$}
&
\multicolumn{2}{c}{$M_A=4191~\MeV$}
\\
Observable & Full calculation & $|\bm P_f|^3$ reference & Full calculation & $|\bm P_f|^3$ reference \\
\hline
$R_n$ & 0.2529 & 1.0917 & 0.0452 & 1.1771 \\
$R_s$ & 1.6298 & 0.3612 & 3.6347 & 0.5887 \\
$A_{sn}$ & 0.731 & -0.503 & 0.975 & -0.333 \\
\end{tabular}
\end{ruledtabular}
\end{table}

\begin{figure}[t]
\centering
\includegraphics[width=0.88\textwidth]{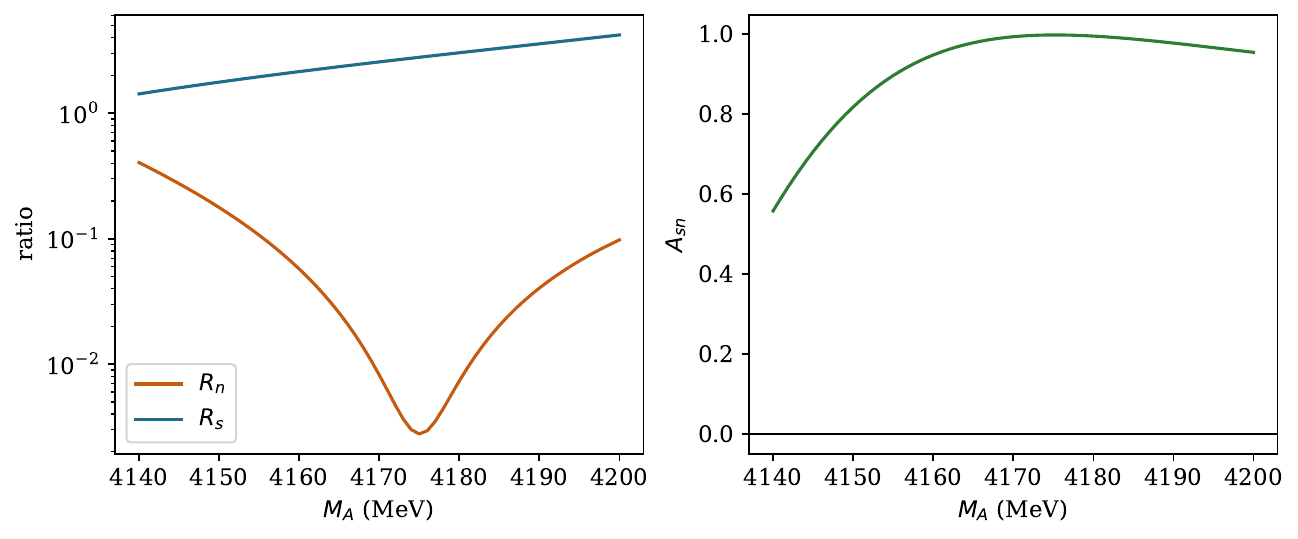}
\caption{
Mass dependence of the same-flavor spin-replacement ratios $R_n$ and
$R_s$ (left) and the bounded asymmetry $A_{sn}$ (right). For the fixed
pure-$2D$ wave function, the ordering $R_n<R_s$ persists across the
sampled mass range.
}
\label{fig:ratios_mass_scan}
\end{figure}

The external-mass dependence is shown in Fig.~\ref{fig:ratios_mass_scan}.
Across $4140$--$4200~\MeV$, the fixed pure-$2D$ wave function gives
$R_n<R_s$. The charge-summed
$D\bar D^*+\mathrm{c.c.}$ width reaches a sampled minimum of about
$0.02367~\MeV$ at $4175~\MeV$. This minimum results from the incoherent
sum of charge sectors: the representative neutral and charged amplitudes
have separate near-zeros around $4171$ and $4179~\MeV$, respectively,
rather than cancelling each other. The increase of $A_{sn}$ toward unity
is therefore driven by the suppression of $R_n$ while $R_s$ remains finite.
The microscopic origin of this suppression is traced to the
channel-dependent overlap structure of the $2D$ wave function in the
following subsection.

After removing the common $P$-wave threshold factor, the resulting
$2\times2$ flavor--spin response also remains strongly non-factorizing.
The corresponding matrix diagnostic is presented in
Appendix~\ref{sec:nonfactorization}.

\subsection{Charge-resolved $D\bar D^*$ zeros and recoil-space scaling}
\label{sec:recoil_zeros}
The charge-summed minimum can be resolved at the projected-amplitude level. For each representative nonstrange $PV$ charge sector, we follow the nonzero $M_{12}$ component through its sign change as a function of the external mass. On the central integration grid the two zeros occur near
\begin{align}
D^0\bar D^{*0}:&\quad M_0\simeq4171~\MeV,\qquad P_{f,0}\simeq0.7753~\GeV,\nonumber\ \\
D^+D^{*-}:&\quad M_+\simeq4179~\MeV,\qquad P_{f,+}\simeq0.7754~\GeV.
\label{eq:charge_resolved_zeros}
\end{align}
The external masses are separated by about $7.5~\MeV$, while the recoil momenta differ by only about $7\times10^{-5}~\GeV$. Thus the structurally relevant observation is the near-common recoil condition, $P_f\simeq0.775~\GeV$: the charged and neutral threshold mappings place essentially the same dynamical cancellation condition at different external masses. The more detailed interpolation and grid-stability information is given in Appendix~\ref{app:limits}. The recoil momentum $P_f$ is not the radial integration variable $q$ and should not be identified directly with the Salpeter-component zero near $q\simeq0.90~\GeV$. The result concerns projected decay-vertex amplitudes; backgrounds, neighboring poles, and rescattering can shift or fill a visible experimental dip.
\begin{figure}[htbp]
\centering
\includegraphics[width=0.86\textwidth]{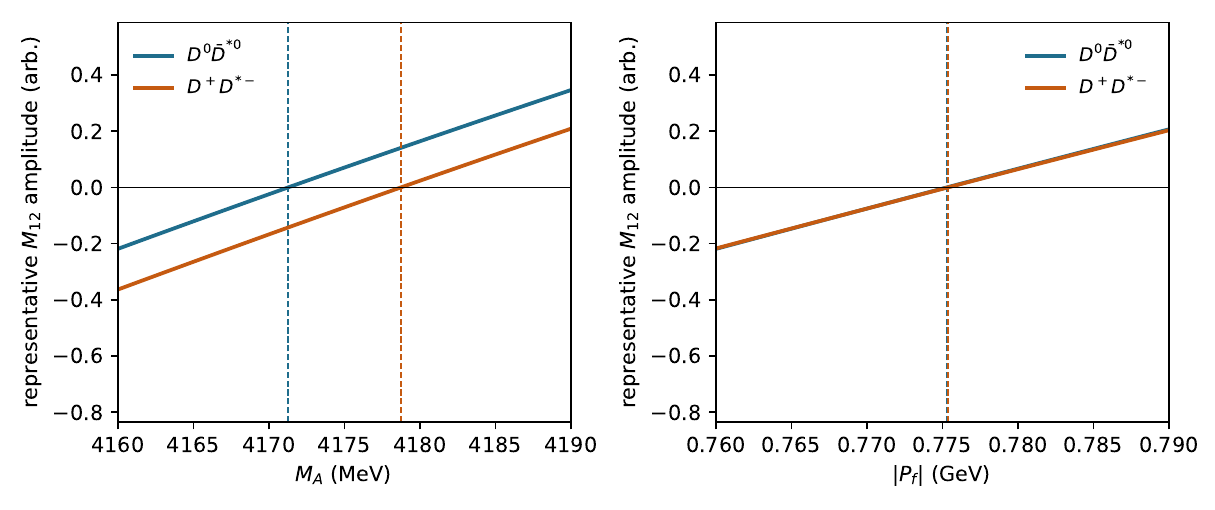}
\caption{Charge-resolved $M_{12}$ amplitudes in the nonstrange $PV$ sector. Left: sign changes as functions of the external mass. Right: the same amplitudes plotted against recoil momentum. The neutral and charged zeros occur at separated masses but map to a nearly common recoil condition, $P_f\simeq0.775~\GeV$.}
\label{fig:recoil_zero}
\end{figure}
To test whether this agreement is accidental, we introduce a continuous external-mass interpolation $m_D(x)=m_{D^0}+x(m_{D^+}-m_{D^0})$ and $m_{D^*}(x)=m_{D^{*0}}+x(m_{D^{*+}}-m_{D^{*0}})$, with $0\leq x\leq1$. This is a kinematic continuation only: the neutral-sector radial inputs and all model parameters are kept fixed, so intermediate $x$ values do not represent physical mesons or self-consistent bound-state solutions. For every $x$, we locate the zero of the same projected $M_{12}$ amplitude.
The interpolation moves the external-mass zero from $4171.28$ to $4177.78~\MeV$, whereas its recoil momentum changes only from $0.77530$ to $0.77407~\GeV$. The relative recoil variation is $0.16\%$, far smaller than the several-MeV shift in the external-mass mapping. The physical charged endpoint, whose final-state wave functions and constituent light mass also differ, is displayed separately in Fig.~\ref{fig:recoil_zero}. Within the present fixed-wave-function positive-energy Salpeter realization, the interpolation therefore supports a recoil-controlled cancellation condition: the channel kernel is tuned by recoil to cancel the weighted overlap with the complete multi-component initial-state wave function. This is not an identification of $P_f$ with the internal radial node near $q\simeq0.90~\GeV$.
The full interpolation trajectory is shown in Appendix~\ref{app:kinematic_interpolation}.

The interpolation test can be strengthened beyond the positions of isolated zeros. Since all channels have lowest relative orbital angular momentum $L=1$, we define the threshold-reduced projected amplitude $F_x(P_f)=\mathcal M_{12,x}(P_f)/P_f$. This is a diagnostic of the calculated amplitude, not a model-independent physical form factor. For the eleven kinematic paths $x=0,0.1,\ldots,1$, we evaluate both the raw $\mathcal M_{12,x}$ and $F_x$ at common recoil values from $0.750$ to $0.800~\GeV$, using the same production integration grid as the baseline calculation.

\begin{figure}[htbp]
\centering
\includegraphics[width=0.92\textwidth]{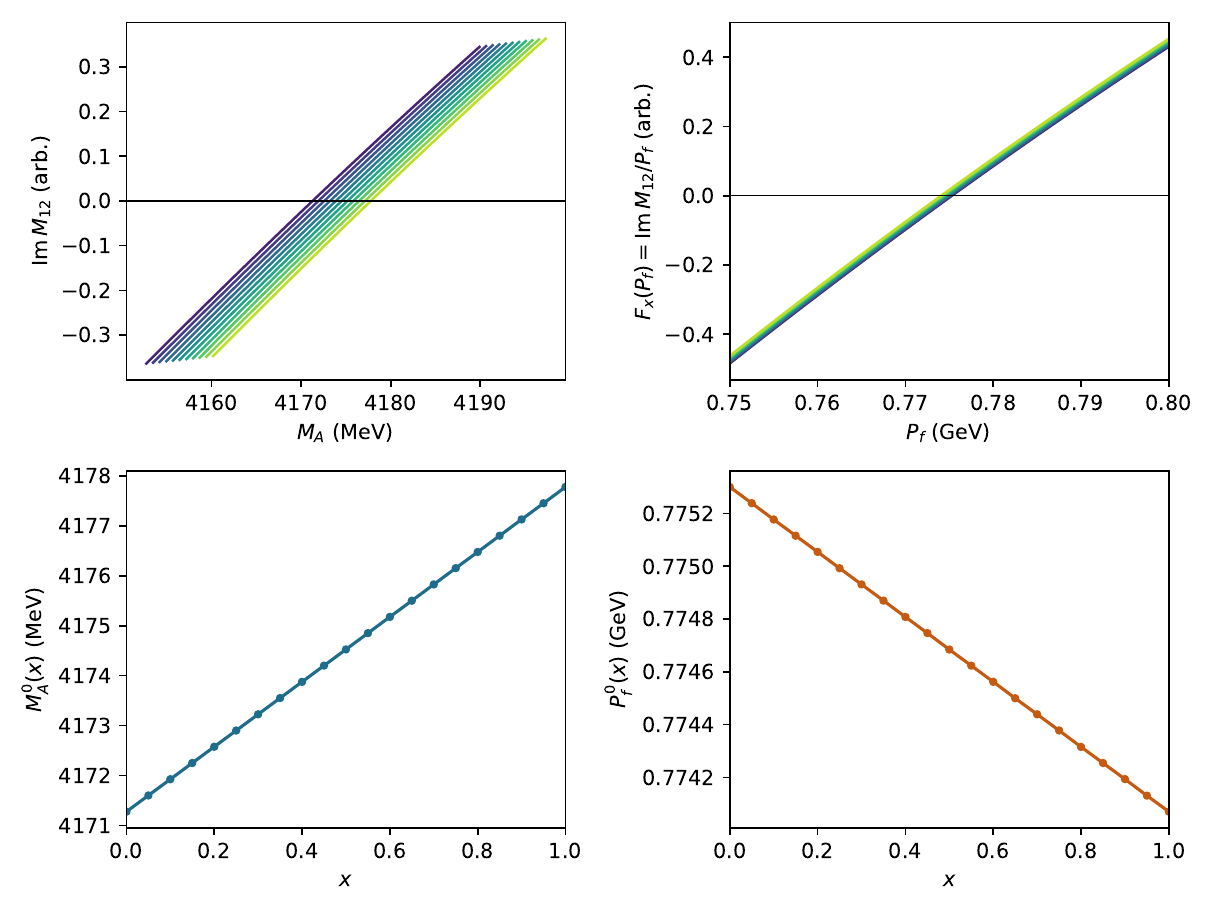}
\caption{Recoil-space scaling test for the kinematic charged-neutral continuation. Upper left: raw projected amplitudes vs $M_A$. Upper right: the same amplitudes divided by $P_f$ vs recoil momentum. The eleven curves in the upper panels correspond to $x=0,0.1,\ldots,1$, with the color progression following increasing $x$. Lower panels: interpolated zero mass and zero recoil momentum.}
\label{fig:recoil_scaling}
\end{figure}

Fig.~\ref{fig:recoil_scaling} displays the resulting mass-space separation and recoil-space behavior. The zero locations separate visibly when plotted against $M_A$, but the threshold-reduced curves form a narrow local family when plotted against $P_f$. To quantify this statement without dividing by the amplitude near its zero, let $\bar F(P_f)$ be the average over the eleven paths and define $\Delta_{\rm coll}=\sqrt{\langle[F_x(P_f)-\bar F(P_f)]^2\rangle_{x,P_f}}/\max_{P_f}|\bar F(P_f)|$. In the indicated recoil window, we obtain $\Delta_{\rm coll}=0.0150$, while the interpolated zero spread is $\Delta P_f^0=0.00123~\GeV$. These are internal numerical diagnostics rather than statistical uncertainties. Within the fixed-input continuation, the projected amplitude is therefore organized more naturally by recoil than by its external mass coordinate.

The numerical robustness of this recoil organization is examined in
Appendix~\ref{app:limits}. A self-consistent heavy-light continuation
preserves the common recoil condition, and an expanded-support calculation
shows that the result is insensitive to the available radial-grid endpoint
within the present implementation.

\section{Momentum-space origin of the pure-$2D$ channel pattern}
\label{sec:nodal}

In the baseline calculation, the open-charm ratio pattern of a $2D$-dominated $\psi(4160)$ reflects both phase space and channel-dependent overlap with the nodal region of the Bethe--Salpeter wave function. The central dynamical observation is that the four matched final states do not probe the same momentum regions with the same weight, even though they originate from the same initial Salpeter wave function and share the same leading threshold power. The four-channel basis in Eq.~\eqref{eq:channel_basis} turns this observation into a two-axis diagnostic: horizontal comparisons probe final-state spin replacement, while vertical comparisons probe light-flavor replacement.

\subsection{Radial node of the $2D$ Bethe--Salpeter wave function}

The first step is to display the sign-changing behavior of the four independent $2D$ Salpeter components and quantify their relative support around the common nodal region.

\begin{figure}[htbp]
\centering
\includegraphics[width=0.8\textwidth]{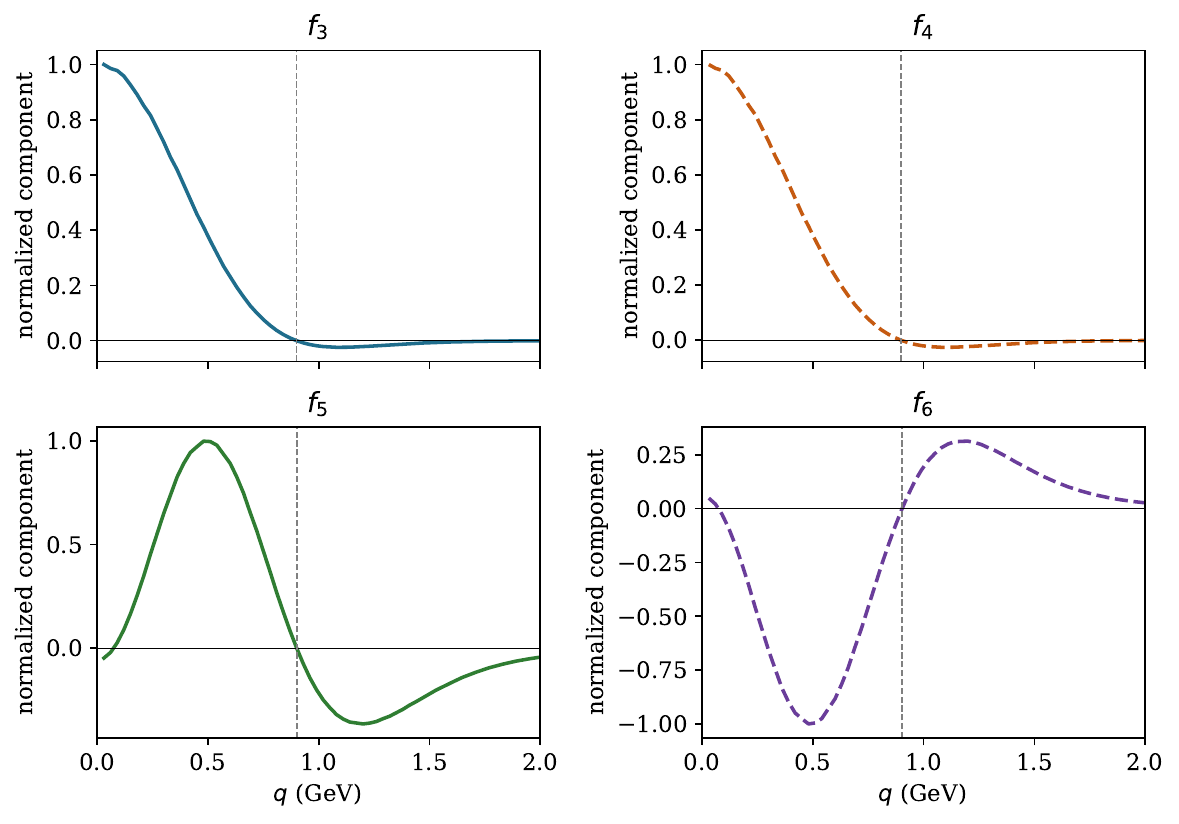}
\caption{Independent Salpeter components of the baseline $2D$ state, each individually normalized to its own maximum absolute value. All four components cross zero near $q\simeq0.90$ GeV, while the post-zero local and integrated support is substantially larger for $f_5$ and $f_6$ than for $f_3$ and $f_4$. The individual normalization displays only the component shapes; their relative nodal support is quantified in Appendix~\ref{app:stability}.}
\label{fig:wavefunction_node}
\end{figure}

The nodal region is at $q\approx0.90~\mathrm{GeV}$. All four components cross zero near this location, but the post-zero support is substantially larger for $f_5$ and $f_6$ than for $f_3$ and $f_4$; the quantitative component audit is collected in Appendix~\ref{app:stability}.

\subsection{Four-channel overlap integrands}

The radial integrands are defined by Eq.~\eqref{eq:radial_integrand_definition}. At tree level the projected amplitudes share a common overall phase, which is removed so that the displayed $I_{a\lambda}(q)$ are real signed functions. For a $PP$ channel there is one projected $P$-wave amplitude in the present convention. For each $PV$ channel we display the representative $M_{12}$ component; $M_{21}=-M_{12}$ in the adopted convention, and the orthogonal polarizations and charge-conjugate modes are summed as described in Appendix~\ref{app:decay_conventions}.

\begin{figure}[htbp]
\centering
\includegraphics[width=0.85\textwidth]{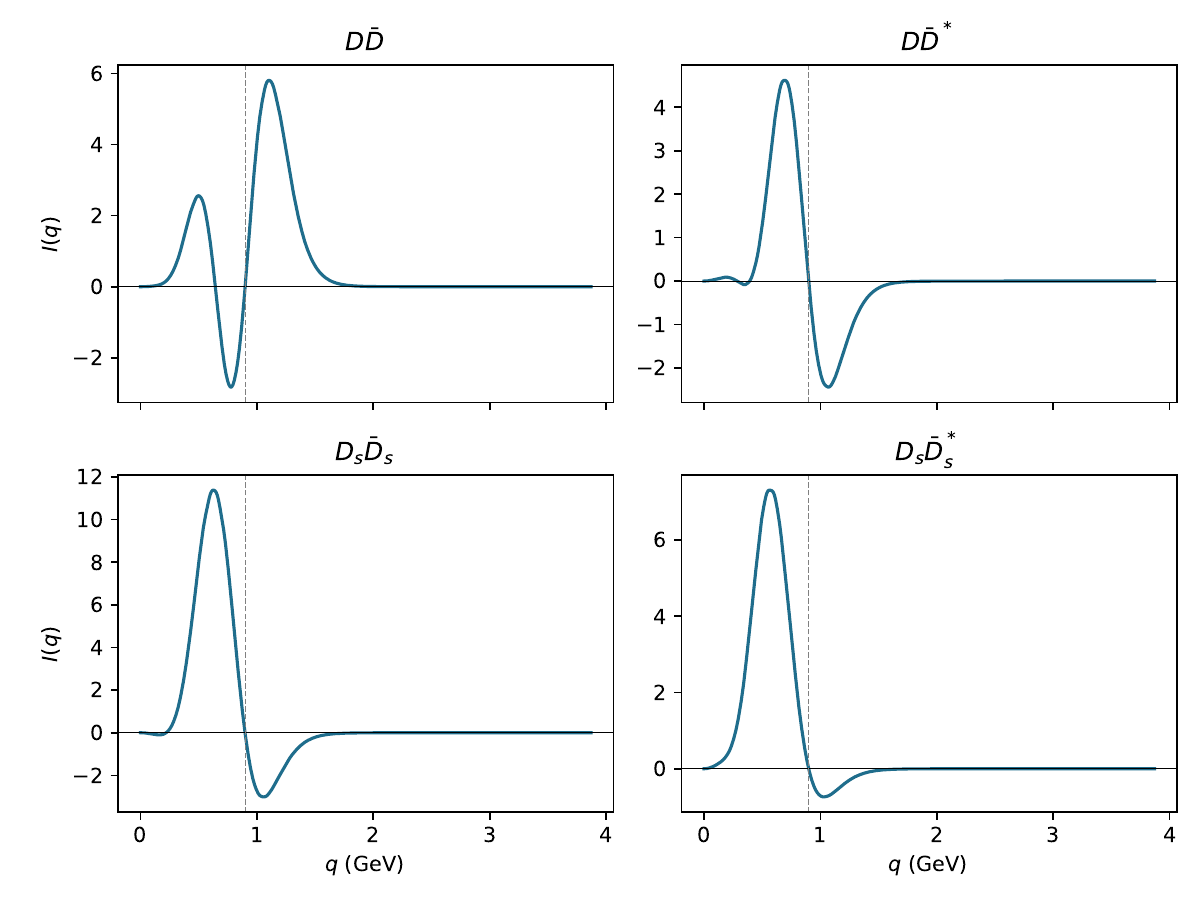}
\caption{Four-channel radial overlap integrands arranged in the $2\times2$ flavor-spin basis at $M_A=4146~\MeV$. The vertical dashed line marks the $2D$ node at $q\approx0.90~\GeV$. The different positive and negative weights illustrate channel-dependent filtering of the common nodal region.}
\label{fig:integrands}
\end{figure}

Figure~\ref{fig:integrands} is the amplitude-level origin of the matched width-ratio pattern: each final state weights the positive and negative sides of the same nodal region differently. The $PP$ integrands and the plotted $PV$ polarization components all change sign near $q\approx0.90~\mathrm{GeV}$. The corresponding $4146~\MeV$ cancellation measures for representative neutral and strange components are listed in Table~\ref{tab:cancellation_indices}.

\subsection{Cumulative amplitudes and cancellation strength}

For the representative projected integrand selected for channel $i$, we use $I_i(q)$ as shorthand for the corresponding $I_{a\lambda}(q)$ defined in Eq.~\eqref{eq:radial_integrand_definition}; for a $PV$ channel this refers to the stated charge sector and polarization component. To quantify its signed overlap without assigning a unique zero to the full integrand, we define
\begin{equation}
P_i=\int_{I_i>0}dq\,I_i(q),\qquad N_i=-\int_{I_i<0}dq\,I_i(q),
\end{equation}
and the normalized cancellation index
\begin{equation}
C_i=1-\frac{|P_i-N_i|}{P_i+N_i}.
\label{eq:cancellation_index}
\end{equation}
A value $C_i\simeq1$ signals a strong cancellation between positive and negative momentum regions. The areas $P_i$ and $N_i$ carry the arbitrary amplitude units inherited from the numerical wave-function normalization; only the dimensionless $C_i$ and ratios formed within the same projected amplitude are compared. These quantities belong to representative projected amplitudes and cannot be assigned directly to charge-summed physical channels, whose charges and polarizations are added incoherently. The most direct matched comparison is $C_{D\bar D^*}=0.7617$ versus $C_{D_s\bar D_s^*}=0.1554$: the representative nonstrange $PV$ amplitude is far more cancellation dominated than its strange counterpart. The remaining channel values are listed in Table~\ref{tab:cancellation_indices}. The plotted kernels and cumulative amplitudes use an independently evaluated uniform radial mesh with $\Delta q=0.005~\GeV$; no post-processing smoothing is applied. The cumulative amplitudes $A_i(Q)=\int_0^Qdq\,I_i(q)$ are shown in Appendix~\ref{app:stability}.

\begin{table}[t]
\caption{Representative projected-amplitude cancellation diagnostics at $M_A=4146~\MeV$. The quantities $P_i$ and $N_i$ are the positive and negative areas of one signed radial integrand in arbitrary amplitude units, and $C_i$ measures its normalized cancellation. For each $PV+\mathrm{c.c.}$ mode, the table reports the $M_{12}$ component; $M_{21}$ has the same $C_i$ under the conventions used here. These indices are not indices of the charge-summed physical channels, which are incoherent sums over charges and polarizations as in Eq.~\eqref{eq:pv_incoherent_sum}.}
\label{tab:cancellation_indices}
\begin{ruledtabular}
\begin{tabular}{lcccc}
Channel & $P_i$ (arb.) & $N_i$ (arb.) & $|P_i-N_i|$ (arb.) & $C_i$ \\
\hline
$D\bar D$ & 2.6343 & 0.4595 & 2.1748 & 0.2970 \\
$D\bar D^*+\mathrm{c.c.}$ & 1.2907 & 0.7940 & 0.4967 & 0.7617 \\
$D_s\bar D_s$ & 3.8821 & 0.9956 & 2.8865 & 0.4082 \\
$D_s\bar D_s^*+\mathrm{c.c.}$ & 2.6765 & 0.2254 & 2.4511 & 0.1554 \\
\end{tabular}
\end{ruledtabular}
\end{table}

\subsection{Flavor--spin response and stress tests}

The four-channel basis provides two complementary replacements: $PP\to PV$ at fixed light flavor and nonstrange $\to$ strange at fixed spin type. The primary observables $R_n$, $R_s$, and $A_{sn}$ summarize the correlated response, while the signed integrands and cancellation indices identify its momentum-space origin.

The radial node alone does not specify the external position of a decay minimum. In the present calculation a zero of a projected amplitude is the recoil-controlled cancellation condition $\mathcal M_i(P_f^0)=g_q\sum_{\alpha=3,4,5,6}\int dq\,K_{i\alpha}(q;P_f^0)f_\alpha(q)=0$, where the kernels include the recoil shift, final-state wave functions, spin projection, and all constrained relativistic components. Thus the internal sign-changing region near $q\simeq0.90~\GeV$ provides the momentum-space environment for cancellation, but it is neither identical to nor sufficient by itself to determine the external scale $P_f^0$.

The sign-removal and node-position tests confirm that the cancellation depends on the signed multicomponent structure of the calculated $2D$ wave function. The restricted $f_5$--$f_6$ deformation preserves the baseline ordering over the tested range, whereas sufficiently large all-component deformations can reverse it. These tests diagnose model sensitivity rather than define a statistical uncertainty; full results are given in Appendix~\ref{app:stability}.

\section{Coherent $3S$--$2D$ structure of the upper state}
\label{sec:mixing_results}

The preceding results resolve the channel kernels acting on the $2D$ basis wave function.  The nearby $3S$ solution must be combined with the same kernels at the amplitude level when the upper state is organized through Eq.~\eqref{eq:psi4160_admixture}.  For a charge-summed channel $i$, let $\mathcal I_{a\lambda}^{(2D)}$ and $\mathcal I_{a\lambda}^{(3S)}$ be the corresponding overlap integrals, evaluated at the same external mass.  In the common real convention of the instantaneous calculation, define
\begin{equation}
\begin{aligned}
D_i&\equiv g_q^2\sum_{a,\lambda}\Phi_{a\lambda}\left|\mathcal I_{a\lambda}^{(2D)}\right|^2=\Gamma_i(2D),\\
S_i&\equiv g_q^2\sum_{a,\lambda}\Phi_{a\lambda}\left|\mathcal I_{a\lambda}^{(3S)}\right|^2=\Gamma_i(3S),\\
X_i&\equiv g_q^2\sum_{a,\lambda}\Phi_{a\lambda}\,
\operatorname{Re}\!\left[\mathcal I_{a\lambda}^{(2D)*}\mathcal I_{a\lambda}^{(3S)}\right].
\end{aligned}
\label{eq:interference_definitions}
\end{equation}
The real two-state branch of Eq.~\eqref{eq:psi4160_admixture} then gives
\begin{equation}
\Gamma_i(\theta,\eta)=D_i\cos^2\theta+S_i\sin^2\theta
+2\eta X_i\sin\theta\cos\theta .
\label{eq:coherent_mixing_width}
\end{equation}
Thus the pure-$2D$ quantities reported above are not discarded: they are the $D_i$ terms whose channel-dependent filtering is reweighted by coherent $3S$--$2D$ interference.  The mass organization fixes only $|\theta|$ within the effective ansatz of Eq.~\eqref{eq:effective_mass_matrix}; it does not select $\eta$.  Moreover, a rephasing of either basis state reverses the signs of $\eta$ and $X_i$ together.  The relevant statement is therefore the channel pattern of the interference in one fixed wave-function convention, rather than an invariant sign assigned to $X_i$ alone.

\begin{table}[t]
\caption{Charge-summed $2D$--$3S$ interference bilinears in MeV at $M_A=4146~\MeV$. The coefficients are defined by Eqs.~\eqref{eq:interference_definitions} and~\eqref{eq:coherent_mixing_width} in the common real convention of the instantaneous calculation. The three columns are rounded independently, so near-saturated interference bilinears can differ from the exact Cauchy bound at the last displayed digit.}
\label{tab:interference_bilinears}
\begin{ruledtabular}
\begin{tabular}{lrrr}
Channel & $D_i$ & $S_i$ & $X_i$ \\
\hline
$D\bar D$ & 5.9731 & 5.7427 & $-5.8567$ \\
$D\bar D^*+\mathrm{c.c.}$ & 1.5106 & 8.4168 & 3.5639 \\
$D_s\bar D_s$ & 4.2023 & 3.8017 & $-3.9970$ \\
$D_s\bar D_s^*+\mathrm{c.c.}$ & 6.8487 & 24.7612 & 13.0224 \\
\end{tabular}
\end{ruledtabular}
\end{table}

For a fixed positive value of $|\theta|$, the choices $\eta=\pm1$ are the two real interference branches. The interference bilinears are negative in the two $PP$ channels and positive in the two $PV$ channels in the supplied convention. Consequently, the $\eta=+1$ branch suppresses $PP$ while enhancing $PV$, most strongly in the strange $PV$ channel, whereas the $\eta=-1$ branch follows the opposite real-angle trajectory. The mass organization selects neither branch. They are therefore conditional coherent decay patterns in a fixed wave-function convention, to be constrained by mixing dynamics or by a common-pole amplitude analysis rather than by the mass spectrum alone.

For illustration, inserting the mass-organized $|\theta|=28.1^\circ$ into the fixed bilinears gives the two branch patterns in Table~\ref{tab:mass_organized_branches}. The near-vanishing $PV$ widths in the $\eta=-1$ column make its derived ratios particularly interference sensitive. These values are algebraic realizations of the effective two-state organization, not a new dynamical determination of the mixing interaction.

\begin{table}[H]
\caption{Conditional charge-summed upper-state patterns at $M_A=4146~\MeV$ for the mass-organized $|\theta|=28.1^\circ$. The two columns use the bilinears of Table~\ref{tab:interference_bilinears} in the fixed real convention. Widths are in MeV; the ratios are formed from the displayed branch widths before rounding. The mass spectrum alone does not select either column.}
\label{tab:mass_organized_branches}
\begin{ruledtabular}
\begin{tabular}{lrr}
Quantity & $\eta=+1$ & $\eta=-1$ \\
\hline
$\Gamma_{D\bar D}$ & 1.06 & 10.79 \\
$\Gamma_{D\bar D^*+\mathrm{c.c.}}$ & 6.00 & 0.081 \\
$\Gamma_{D_s\bar D_s}$ & 0.79 & 7.43 \\
$\Gamma_{D_s\bar D_s^*+\mathrm{c.c.}}$ & 21.64 & 0.0012 \\
$R_n$ & 5.69 & 0.0075 \\
$R_s$ & 27.33 & 0.00016 \\
$A_{sn}$ & 0.655 & $-0.957$ \\
\end{tabular}
\end{ruledtabular}
\end{table}

Fig.~\ref{fig:theta_width_scan} shows the corresponding real-angle scan;
the detailed widths are tabulated in Appendix~\ref{app:stability}.  The narrow destructive regions, where a same-flavor ratio becomes small, are interference-sensitive and should not be treated as a mass-spectrum prediction.  The broader conclusion is instead structural: the bare-spectrum tension, the component-resolved $2D$ channel filters, and the final decay pattern are linked through the same pair of BS solutions and their coherent amplitudes.  Appendix~\ref{app:stability} records the additional sensitivity to a general relative phase, which is outside the real effective two-state organization.

\begin{figure}[htbp]
\centering
\includegraphics[width=0.95\textwidth]{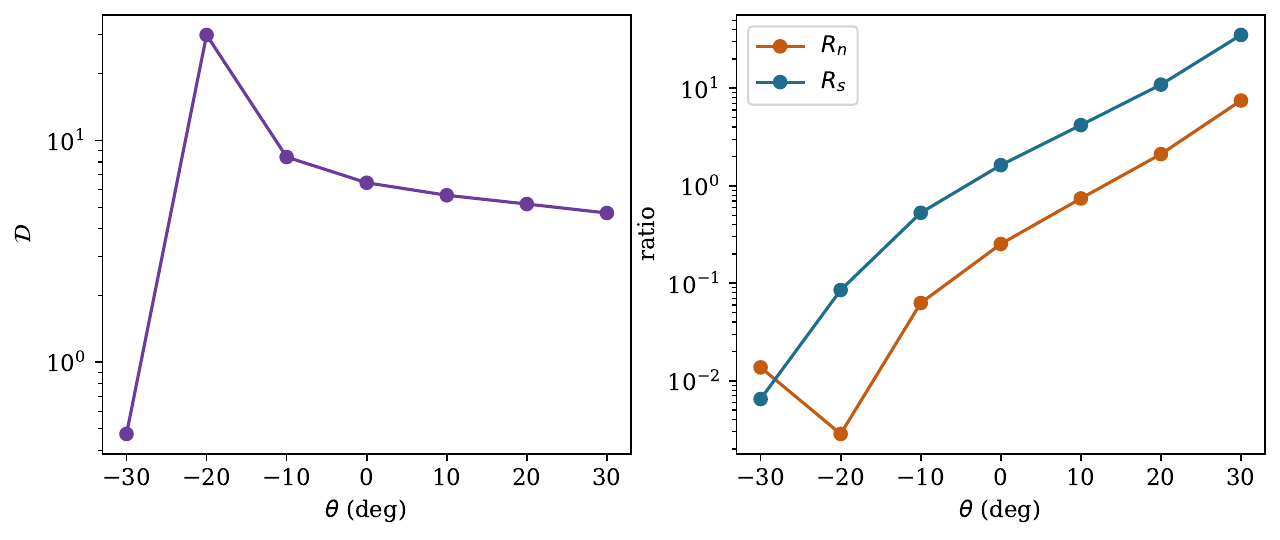}
\caption{Real-admixture scan at $M_A=4146~\MeV$ using the supplied relative phase. Left: the double ratio $\mathcal D$ is displayed on a logarithmic scale. Right: the same-flavor ratios.}
\label{fig:theta_width_scan}
\end{figure}

\section{Discussion}
\label{sec:discussion}

\subsection{Physical interpretation and model boundaries}

The spectrum and the decay mechanism are two aspects of one construction.  The BS equation supplies nearby bare $3S$ and $2D$ solutions together with their multi-component wave functions.  Their opposite displacement from the two observed vector masses motivates the effective two-state organization of Eq.~\eqref{eq:effective_mass_matrix}; the same basis amplitudes then enter the open-charm kernels coherently.  At the component level, the full $2D$ calculation reverses the strange-versus-nonstrange ordering of the common $P$-wave threshold reference, the signed radial integrands show strongly channel-dependent cancellation, and the charge-resolved $D\bar D^*$ zeros align near a common recoil condition.  At the upper-state level, the $3S$ amplitude reweights these component-resolved structures through channel-dependent interference.  These are not independent mechanisms but successive consequences of the same BS basis and channel-resolved overlaps.

The effective mass matrix is a phenomenological organization of the spectrum; it is not a coupled-channel dynamical calculation of the $S$--$D$ mixing matrix element. In particular, it determines a representative $|\theta|$ but neither the relative decay branch nor a general complex phase.  The pure-$2D$ recoil zero is correspondingly a basis-component benchmark; a physical mixed-state zero would require a coherent scan with a specified mixing dynamics.  The central decay prediction is therefore the coherent channel pattern and its pole residues, rather than a universal zero in a measured cross section.  A detailed discussion of the model boundaries, including the fixed-wave-function approximation and the remaining numerical limitations, is given in Appendix~\ref{app:limits}.

\subsection{Recoil-controlled cancellation and non-factorizing response}

The charged-neutral continuation strengthens the interpretation of the nonstrange $PV$ minimum. Varying the external masses changes the threshold mapping from $M_A$ to $P_f$, and hence moves the mass at which the projected overlap cancels. The local curve-collapse test shows that this is not only a coincidence of two endpoints: the threshold-reduced projected amplitudes form a narrow recoil-space family, while the sign-removal counterfactual eliminates the zero in the tested recoil interval. In this sense, the zero is naturally interpreted as a recoil-controlled cancellation condition, or equivalently a weighted orthogonality of the channel kernel and the complete initial Salpeter wave function within the specified fixed-input realization. It is not a model-independent orthogonality relation and does not imply that an observed production amplitude must vanish.

The threshold-reduced matrix diagnostic provides a complementary statement. The large positive $\mathcal N_{FS}$ shows that, after the common $P$-wave barrier is removed, the effects of light-flavor and final-state-spin replacement do not factorize. The four open-charm modes may therefore be viewed as channel-resolved, partial momentum-space probes of the parent state: each applies a distinct recoil- and spin-dependent kernel to the same multi-component wave function. This usage is model dependent and does not constitute direct wave-function reconstruction.

\subsection{Relation to earlier charmonium decay calculations}

Our previous $\psi(4040)$ analysis identified strong cancellation in one suppressed channel and followed its mass evolution~\cite{Wan2026}. The present work goes beyond that one-channel mechanism by constructing a matched two-axis flavor--spin basis with common threshold power and by identifying a correlated recoil-space structure. Earlier nonrelativistic $^3P_0$ calculations also established strong node sensitivity in higher-charmonium widths~\cite{Barnes2005,Gui2018}; the comparison here instead emphasizes the correlated response of matched channels. The baseline value $R_n\simeq0.253$ at $4146~\MeV$ lies between the two charge-summed results reconstructed from Ref.~\cite{Gui2018}, while the Barnes--Godfrey--Swanson calculation gives a stronger suppression at a nearby mass~\cite{Barnes2005}. Because the calculations use different masses, wave functions, and pair-creation conventions, these numerical comparisons are illustrative.

\subsection{Experimental implications}

Existing open-charm measurements cover much of the required kinematic region: BESIII has measured $e^+e^-\to D\bar D$, $D^{*+}D^-$, and $D_s^+D_s^-$ across the $\psi(4160)$ region~\cite{BESIII_DD_2024,BESIII_DstarD_2022,BESIII_DsDs_2024}, while Belle ISR data provide complementary strange open-charm channels~\cite{Belle_DsDsstar_2011}. A production cross section cannot be identified directly with a partial width because nonresonant production, neighboring vector states, and channel-dependent interference can modify the visible line shapes. A common amplitude analysis may write the contribution to channel $i$ near the $\psi(4160)$ pole as
\begin{equation}
\mathcal A_i(s)=\frac{c_{\rm prod}\,g_i}{s-s_p}+\mathcal A_i^{\rm nr}(s),
\qquad r_i=c_{\rm prod}\,g_i,
\label{eq:pole_residue_mapping}
\end{equation}
so that, under one common pole and normalization convention,
\begin{equation}
\frac{|r_i|^2}{|r_j|^2}=\frac{|g_i|^2}{|g_j|^2}.
\label{eq:residue_ratio}
\end{equation}
Consistent two-body kinematic factors then allow $R_n$, $R_s$, and $A_{sn}$ to be constructed from the fitted residues.

The pure-$2D$ calculation supplies a structural benchmark: it gives $R_s>R_n$, equivalently $A_{sn}>0$, whereas the purely kinematic threshold reference has the opposite ordering at the central benchmark. A common-pole fit accesses the residues of the physical mixed state, not a separately measurable pure-$2D$ residue. It can nevertheless test whether the physical channel pattern is consistent with the coherent reweighting of this benchmark. A more differential diagnostic is the charge-resolved nonstrange $PV$ suppression: the $2D$ component organizes the neutral and charged projected zeros around a common recoil scale, $P_f\sim0.775~\GeV$, rather than a common external mass. A recoil-correlated suppression pattern in a unified amplitude fit would provide a concrete test of the nodal-filtering interpretation; no visible cross-section zero is required.

A coherent amplitude analysis need not reveal a visible cross-section zero. The more informative signature is an energy-dependent evolution of the resonant contribution relative to coherent backgrounds across the recoil-controlled cancellation region. The tree-level projected sign reversal motivates this fit target, but neither a universal phase jump nor a directly observable sign is implied once rescattering and background phases are present.

\section{Summary}
\label{sec:summary}

We have demonstrated that open-charm decays can provide a channel-resolved probe of the momentum-space structure of heavy quarkonium wave functions. By following a common Bethe--Salpeter basis from the bare spectrum to coherent decay amplitudes, we establish a direct connection between the internal sign-changing structure of the initial state and experimentally accessible open-charm decay patterns. In the $\psi(4040)$--$\psi(4160)$ system, the BS equation yields nearby $3\,^3S_1$ and $2\,^3D_1$ configurations at 4051 and 4110 MeV, respectively. Their opposite displacements relative to the physical vector states motivate an effective $3S$--$2D$ level-repulsion organization, while the corresponding decay amplitudes retain the full multi-component momentum-space information of the BS wave functions.

The central result is that different open-charm channels do not simply measure decay probabilities of the same parent state, but act as distinct momentum-space filters of its internal structure. Using the matched channels $D\bar D$, $D\bar D^*+{\rm c.c.}$, $D_s\bar D_s$, and $D_s\bar D_s^*+{\rm c.c.}$, which share the same leading $P$-wave threshold behavior, we identify a channel-dependent nodal filtering mechanism associated with the $2D$ Salpeter wave function. At $M_A=4146$ MeV, the full calculation gives a nonstrange $PV/PP$ ratio of 0.2529 and a strange-sector ratio of 1.6298, whereas the corresponding threshold-only expectation exhibits the opposite ordering. This reversal demonstrates that the observed channel hierarchy is controlled by dynamical overlap effects rather than phase space alone.

The momentum-space origin of this behavior is revealed by the signed decay integrands. The independent $2D$ Salpeter components change sign near $q\simeq0.90$ GeV, while different final-state kernels weight the two sides of this nodal region with different strengths. The resulting cancellation pattern is therefore channel dependent: the representative cancellation indices are $C_{D\bar D^*}=0.7617$ and $C_{D_s\bar D_s^*}=0.1554$, showing that the nonstrange $PV$ amplitude is much more strongly suppressed by destructive overlap than its strange counterpart. Thus, the decay pattern encodes information about the momentum-space node rather than only the available phase space.

A further characteristic signature appears in the charge-resolved nonstrange $PV$ amplitudes. Although the neutral and charged channels vanish at different external masses near 4.17 and 4.18 GeV, respectively, they correspond to an almost identical recoil momentum, $P_f\simeq0.775$ GeV. The continuous interpolation confirms that the suppression is organized by the recoil-controlled overlap condition rather than by a particular external mass value. This behavior reflects the full shifted overlap between the relativistic multi-component initial-state wave function and the channel-dependent decay kernels, rather than a direct identification with the internal radial node.

Finally, the coherent combination of the $3S$ and $2D$ amplitudes shows how the physical upper state inherits and reweights the component-resolved filtering pattern through interference. The two real interference branches generate distinct $PP$--$PV$ decay structures, while their relative realization requires dynamical information beyond the mass spectrum. These results provide experimentally testable signatures through a common-pole amplitude analysis of open-charm data, including the residue hierarchy among channels, recoil-correlated suppression structures, and the coherent flavor-spin reorganization induced by $3S$--$2D$ interference.

Together, the present study shows that strong decays can serve not only as probes of hadron spectroscopy but also as momentum-space microscopes of confined quark dynamics.

\section*{Acknowledgments}
This work was supported by NSFC Grants No.~12575106 and 12147214, and by the Liaoning Province Fundamental Research Fund No.~LJ212410165019. The authors used OpenAI ChatGPT (GPT-5.6) and Aether (v0.72) to assist with manuscript revision, scientific consistency checks, and limited code generation, debugging, plotting, and reproducibility verification. The authors specified the physical assumptions and validation criteria, reviewed all outputs, and take full responsibility for the work.

\appendix

\section{Full Bethe--Salpeter setup}
\label{app:bs}
The main text uses only the ingredients directly required for the nodal-filtering argument. For completeness, the standard instantaneous Bethe--Salpeter reduction and the vector-state decomposition used in the numerical calculation are recorded here.

The instantaneous Bethe--Salpeter equation~\cite{Salpeter1952,SalpeterOrigins,Chang2005,Chang2004,Wang2006} is used to obtain the charmonium wave functions and the corresponding bare masses. This section first summarizes the basic definitions of the bound-state calculation and then uses the resulting spectrum to define the spectroscopic setting. The same Salpeter wave functions will later enter the open-charm decay amplitudes, so the notation introduced here also fixes the convention used in the decay analysis.

For a quark-antiquark bound state with total momentum $P$ and mass $M$, the quark and antiquark momenta are written as
\begin{equation}
p_1=\alpha_1P+q,\qquad
p_2=\alpha_2P-q,
\qquad
\alpha_i=\frac{m_i}{m_1+m_2},
\label{eq:bs_momenta}
\end{equation}
where $q$ is the relative momentum. The four-dimensional BS wave function $\chi_P(q)$ satisfies
\begin{equation}
S_1^{-1}(p_1)\chi_P(q)S_2^{-1}(-p_2)
=i\int\frac{d^4k}{(2\pi)^4}V(P;q,k)\chi_P(k),
\label{eq:bs_equation}
\end{equation}
with $S_i$ the constituent-quark propagators and $V$ the interaction kernel. In the instantaneous approximation, the kernel depends only on the components transverse to $P$,
\begin{equation}
V(P;q,k)\simeq V(q_\perp,k_\perp),
\qquad
q_\perp^\mu=q^\mu-\frac{P\cdot q}{M^2}P^\mu .
\label{eq:instantaneous_kernel}
\end{equation}
The corresponding three-dimensional Salpeter wave function is defined as
\begin{equation}
\varphi_P(q_\perp)=i\int\frac{dq_P}{2\pi}\chi_P(q),
\qquad
q_P=\frac{P\cdot q}{M} .
\label{eq:salpeter_wavefunction}
\end{equation}

The quark propagator is decomposed into positive- and negative-energy parts through the projectors
\begin{equation}
\Lambda_i^\pm(q_\perp)=\frac{1}{2\omega_i}
\left[
\frac{\slashed P}{M}\omega_i
\pm J_i\left(m_i+\slashed q_\perp\right)
\right],
\qquad
\omega_i=\sqrt{m_i^2-q_\perp^2},
\label{eq:energy_projectors}
\end{equation}
where $J_1=+1$ for the quark and $J_2=-1$ for the antiquark. This form uses the common relative momentum $q_\perp$ defined by Eq.~\eqref{eq:bs_momenta}; equivalently, with the constituent transverse momentum $p_{i\perp}=J_i q_\perp$, the second term is $\pm(\slashed p_{i\perp}+J_i m_i)$. In the meson rest frame, $q_\perp^2=-\mathbf q^2$ and therefore $\omega_i=\sqrt{m_i^2+\mathbf q^2}$. The projected Salpeter components are
\begin{equation}
\varphi_P^{\pm\pm}(q_\perp)=
\Lambda_1^\pm(q_\perp)\frac{\slashed P}{M}
\varphi_P(q_\perp)\frac{\slashed P}{M}
\Lambda_2^\pm(q_\perp).
\label{eq:projected_wavefunctions}
\end{equation}
After the integration over $q_P$, the Salpeter equation takes the standard coupled form
\begin{align}
(M-\omega_1-\omega_2)\varphi_P^{++}(q_\perp)
&=\Lambda_1^+(q_\perp)\eta_P(q_\perp)\Lambda_2^+(q_\perp),\label{eq:salpeter_pp}\\
(M+\omega_1+\omega_2)\varphi_P^{--}(q_\perp)
&=-\Lambda_1^-(q_\perp)\eta_P(q_\perp)\Lambda_2^-(q_\perp),\label{eq:salpeter_mm}\\
\varphi_P^{+-}(q_\perp)&=0,
\qquad
\varphi_P^{-+}(q_\perp)=0,
\label{eq:salpeter_constraints}
\end{align}
where
\begin{equation}
\eta_P(q_\perp)=\int\frac{d^3k_\perp}{(2\pi)^3}
V(q_\perp,k_\perp)\varphi_P(k_\perp).
\label{eq:eta_definition}
\end{equation}

In the numerical calculation, we use a screened Cornell-type kernel in momentum space,
\begin{align}
V(\mathbf q)&=(2\pi)^3V_s(\mathbf q)
+\gamma^0\otimes\gamma_0(2\pi)^3V_v(\mathbf q),\label{eq:kernel_total}\\
V_s(\mathbf q)&=-\left(\frac{\lambda}{\alpha}+V_0\right)\delta^3(\mathbf q)
+\frac{\lambda}{\pi^2}\frac{1}{(\mathbf q^2+\alpha^2)^2},\label{eq:kernel_scalar}\\
V_v(\mathbf q)&=-\frac{2}{3\pi^2}\frac{\alpha_s(\mathbf q)}{\mathbf q^2+\alpha^2},
\label{eq:kernel_vector}
\end{align}
with the running coupling
\begin{equation}
\alpha_s(\mathbf q)=\frac{12\pi}{27}
\frac{1}{\log\left(a+\mathbf q^2/\Lambda_{\rm QCD}^2\right)} .
\label{eq:running_coupling}
\end{equation}
The complete numerical values of the quark masses and kernel parameters are collected in Appendix~\ref{app:inputs}. Solving Eqs.~\eqref{eq:salpeter_pp}--\eqref{eq:eta_definition} with the screened Cornell kernel in Eqs.~\eqref{eq:kernel_total}--\eqref{eq:running_coupling} gives the vector charmonium spectrum used in the main text.

For the vector charmonium states with $J^{PC}=1^{--}$, the instantaneous Salpeter wave function can be decomposed into eight Lorentz structures,
\begin{align}
\varphi_{1^-}(q_\perp)=&
(q_\perp\cdot\epsilon)\left[
f_1(q_\perp)
+\frac{\slashed P}{M}f_2(q_\perp)
+\frac{\slashed q_\perp}{M}f_3(q_\perp)
+\frac{\slashed P\slashed q_\perp}{M^2}f_4(q_\perp)
\right]
\nonumber\\
&+M\slashed\epsilon f_5(q_\perp)
+\slashed\epsilon\slashed P f_6(q_\perp)
+\left(\slashed q_\perp\slashed\epsilon-q_\perp\cdot\epsilon\right)f_7(q_\perp)
\nonumber\\
&+\frac{1}{M}\left(\slashed P\slashed\epsilon\slashed q_\perp
-\slashed P\,q_\perp\cdot\epsilon\right)f_8(q_\perp),
\label{eq:vector_wavefunction}
\end{align}
where $\epsilon^\mu$ is the polarization vector and the scalar functions $f_i(q_\perp)$ depend only on $q_\perp^2$. The Salpeter constraints $\varphi^{+-}=\varphi^{-+}=0$ reduce the number of independent radial functions from eight to four. In the convention used here, $f_3$, $f_4$, $f_5$, and $f_6$ are taken as independent, while $f_1$, $f_2$, $f_7$, and $f_8$ are determined by the constraint equations. These scalar functions reconstruct the positive-energy component used in the decay amplitudes and are the components plotted in Fig.~\ref{fig:wavefunction_node}. The constraints $\varphi^{+-}=\varphi^{-+}=0$ hold in the Salpeter reduction, whereas neglecting $\varphi^{--}$ in the decay matrix element is an additional approximation. This positive-energy approximation is standard in the Salpeter phenomenology of open-flavor decays~\cite{Chang2005,Wang2006}. A quantitative estimate of the residual effect is not included. The normalization-independent ratios may be less sensitive than absolute widths to a common normalization shift, but channel-dependent corrections remain possible, especially near an amplitude zero.

To specify the numerical convention, the leading positive-energy decay projection evaluates traces of the form~\cite{Wang2013Strong,Wan2026}
\begin{equation}
\mathcal T_{a\lambda}=\operatorname{Tr}\!\left[
\bar\varphi_{B,a}^{++}(q_{1\perp})\,
\varphi_A^{++}(q_\perp)\,
\bar\varphi_{C,a}^{++}(q_{2\perp})\right]_{\lambda}.
\label{eq:salpeter_decay_trace}
\end{equation}
The scalar $^3P_0$ insertion is the unit Dirac matrix for the interaction $\bar\psi_q\psi_q$. The displayed trace is the leading positive-energy instantaneous trace used in the calculation. The subscript $\lambda$ denotes the $PP$ projection or a pair $(r,s)$ of initial- and final-vector polarizations for $PV$. In the rest frame of $A$, label the spectator constituent by $1$, the initial constituent connected to the created pair by $2$, and the created constituent by $3$. With $\vartheta$ the angle between $\bm q$ and $\bm P_f$, the radial arguments used by the code are
\begin{align}
q_{1T}^2&=q^2+\alpha_{13}^2P_f^2-2\alpha_{13}qP_f\cos\vartheta,
&\alpha_{13}&=\frac{m_1}{m_1+m_3},\nonumber\\
q_{2T}^2&=q^2+\alpha_{23}^2P_f^2-2\alpha_{23}qP_f\cos\vartheta,
&\alpha_{23}&=\frac{m_2}{m_2+m_3},
\label{eq:recoil_arguments}
\end{align}
with $P_f=\sqrt{\lambda(M_A^2,M_B^2,M_C^2)}/(2M_A)$ and $\lambda(x,y,z)=x^2+y^2+z^2-2xy-2xz-2yz$. Equations~\eqref{eq:salpeter_decay_trace} and~\eqref{eq:recoil_arguments}, the standard reconstruction cited below, and Table~\ref{tab:charge_conventions} define the numerical projection.

The positive-energy $1^-$ component is reconstructed from the four independent functions $f_3$, $f_4$, $f_5$, and $f_6$ in a coefficient basis $(a,b,c,d,e,f,g,h)$. The coefficient definitions and the algebraic relations imposed by the Salpeter constraints follow the standard instantaneous Bethe--Salpeter conventions of Refs.~\cite{Chang2005,Chang2004,Wang2006}. The same reconstruction is applied to the vector open-charm wave function with its corresponding masses and radial functions; the pseudoscalar projection uses its two independent radial functions and the same positive-energy projector convention. In the decay calculation the tabulated radial wave functions are held fixed, whereas the external initial-state mass is varied in the kinematics and in the positive-energy coefficient reconstruction.

This multi-component structure is important for the present problem. For a radially excited vector charmonium, the node is not a zero of a single nonrelativistic radial wave function, but appears through the combined behavior of the Salpeter components in Eq.~\eqref{eq:vector_wavefunction}. Consequently, the later discussion of nodal filtering will refer to the full channel-dependent overlap integrand constructed from these components, rather than to any one $f_i$ alone.

\section{Decay conventions and channel counting}
\label{app:decay_conventions}

The numerical decay amplitudes are evaluated with the same charge and
polarization conventions used in the main text. For a given charge sector
$a$, the amplitude is first constructed for each projected polarization
component $\lambda$. The charge sum and polarization sum are performed only
at the squared-amplitude level according to Eq.~\eqref{eq:width}.

For the $PV$ channels, the projected polarization amplitude is denoted as
$M_{rs,a}$, where $r$ and $s$ label the initial-vector and final-vector
polarizations, respectively. Choosing the final-state momentum along the
adopted Cartesian axis, only the transverse components
$M_{12,a}$ and $M_{21,a}$ contribute. These two components are summed as
independent polarization amplitudes and do not represent additional charge
multiplicities. Because they correspond to distinct polarization states,
the polarization sum contains their squared magnitudes separately. A
further factor of two adds the physical charge-conjugate channel
$V\bar P$,
\begin{equation}
\Gamma_{PV+\mathrm{c.c.}}=
2\sum_a K_a
\left(|M_{12,a}|^2+|M_{21,a}|^2\right),
\qquad
K_a=\frac{|\bm P_{f,a}|}{8\pi M_A^2(2J_A+1)},
\label{eq:pv_incoherent_sum}
\end{equation}
with no $M_{12}M_{21}^*$ interference term because the two amplitudes
belong to orthogonal polarization configurations. In the adopted phase
convention $M_{21}=-M_{12}$, so they have equal squared magnitudes and the
same cancellation and deformation response; their relative sign has no
effect on the polarization-summed width. Charge conjugation gives the
equal-width partner represented by the explicit prefactor.

The charge-state conventions used in the calculation are summarized in
Table~\ref{tab:charge_conventions}. The multiplicity factors include the
explicit charge-conjugate states but do not include additional polarization
multiplicities.

\begin{table}[t]
\caption{Charge-state and mass conventions used in the decay calculation.
For each representative $P\bar V$ sector, $M_{12}$ and $M_{21}$ denote the
two nonzero entries of the $3\times3$ initial--final vector-polarization
matrix. They are summed at the squared-amplitude level before the explicit
charge-conjugate contribution is included.}
\label{tab:charge_conventions}
\begin{ruledtabular}
\begin{tabular}{lclc}
Channel & States included & Masses (GeV) & Multiplicity \\
\hline
$D\bar D$
&
$D^0\bar D^0,\ D^+D^-$
&
$M_{D^0}=1.865,\ M_{D^+}=1.870$
&
2
\\

$D\bar D^*+\mathrm{c.c.}$
&
$D^0\bar D^{*0},\
D^{*0}\bar D^0,\
D^+D^{*-},\
D^{*+}D^-$
&
$M_{D^{*0}}=2.007,\ M_{D^{*+}}=2.010$
&
4
\\

$D_s\bar D_s$
&
$D_s^+D_s^-$
&
$M_{D_s}=1.968$
&
1
\\

$D_s\bar D_s^*+\mathrm{c.c.}$
&
$D_s^+D_s^{*-},\
D_s^{*+}D_s^-$
&
$M_{D_s^*}=2.112$
&
2
\\
\end{tabular}
\end{ruledtabular}
\end{table}

\subsection{Full decay amplitude expression}
\label{app:full_amplitude}

The explicit positive-energy instantaneous Mandelstam expression used in
the numerical calculation is

\begin{equation}
\mathcal M_{a\lambda}^{(q)}
=
g_q
\int
\frac{d^3\bm q}{(2\pi)^3}
\mathrm{Tr}
\left[
\bar\varphi_{B,a}^{++}(\bm q_B)
\varphi_A^{++}(\bm q)
\bar\varphi_{C,a}^{++}(\bm q_C)
\right]_{\lambda}.
\end{equation}

The recoil-shifted momenta $\bm q_B$ and $\bm q_C$ are determined by the
standard Jacobi transformation of the final-state wave functions. Their
explicit expressions are retained here because they enter the numerical
overlap integrals directly.

The plotted radial integrand in the main text is defined as

\begin{equation}
I_{a\lambda}(q)
=
\frac{q^2}{(2\pi)^3}
\int d\Omega_q
\mathrm{Tr}
\left[
\bar\varphi_{B,a}^{++}(\bm q_B)
\varphi_A^{++}(\bm q)
\bar\varphi_{C,a}^{++}(\bm q_C)
\right]_{\lambda}.
\end{equation}

It should be emphasized that this quantity is not factorized into an
initial radial wave function and a channel kernel. The full recoil,
spin, flavor, and Salpeter-component dependence is retained inside the
overlap.

\section{Numerical inputs and extended decay tables}
\label{app:inputs}
The following tables document the numerical parameters and auxiliary quantities used in the calculation.

\begin{table}[t]
\caption{Input parameters used in the Bethe--Salpeter equation and the relativistic ${}^3P_0$ model. The vertex coefficient $g_q$ is the quantity that directly enters the decay amplitude, while $\gamma_q$ is the equivalent dimensionless convention.}
\label{tab:parameters}
\begin{ruledtabular}
\begin{tabular}{lcl}
Parameter & Value & Role \\
\hline
$m_c$ & $1.620~\GeV$ & charm constituent mass \\
$m_u$ & $0.305~\GeV$ & light constituent mass \\
$m_d$ & $0.311~\GeV$ & light constituent mass \\
$m_s$ & $0.500~\GeV$ & strange constituent mass \\
$\lambda$ & $0.210~\GeV^2$ & confinement strength \\
$\alpha$ & $0.06~\GeV$ & screening parameter \\
$a$ & $e=2.7183$ & running-coupling input \\
$\Lambda_{\rm QCD}$ & $0.27~\GeV$ & running-coupling scale \\
$\gamma_n$ & $0.526\pm0.023$ & equivalent dimensionless value fitted from $\psi(3770)\to D\bar D$ \\
$g_n=2m_n\gamma_n$ & $0.324\pm0.014~\GeV$ & nonstrange vertex used in the amplitude \\
$g_s/g_n$ & $1$ & baseline flavor prescription \\
$\gamma_s=(m_n/m_s)\gamma_n$ & $0.324\pm0.014$ & equivalent strange dimensionless value \\
\end{tabular}
\end{ruledtabular}
\end{table}

\begin{table}[t]
\caption{Values of the constant term $V_0$ used for the meson systems relevant to the present calculation. The $c\bar c$ value is used for the vector charmonium wave functions, while the heavy-light values are used for the final open-charm mesons.}
\label{tab:V0values}
\begin{ruledtabular}
\begin{tabular}{lcc}
System & $J^P$ & $V_0$ (GeV) \\
\hline
$c\bar c$ & $1^-$ & $-0.1756$ \\
$c\bar u$ & $0^-$ & $-0.375$ \\
$c\bar d$ & $0^-$ & $-0.375$ \\
$c\bar s$ & $0^-$ & $-0.432$ \\
$c\bar u$ & $1^-$ & $-0.110$ \\
$c\bar d$ & $1^-$ & $-0.110$ \\
$c\bar s$ & $1^-$ & $-0.167$ \\
\end{tabular}
\end{ruledtabular}
\end{table}

The reduced strengths and strange-pair normalization checks are summarized below.

\begin{table}[t]
\caption{Reduced $P$-wave strengths $\widehat\Gamma_i=\Gamma_i/\sum_{a\in i}|\bm P_{f,a}|^3$ for the four selected charge-summed channels, in units of $\MeV/\GeV^3$. These diagnostic quantities remove the common leading threshold factor and expose residual channel dependence from the decay overlap.}
\label{tab:reduced_widths}
\begin{ruledtabular}
\begin{tabular}{lcc}
Channel & $\widehat\Gamma_i(4146)$ & $\widehat\Gamma_i(4191)$ \\
& (MeV/GeV$^3$) & (MeV/GeV$^3$) \\
\hline
$D\bar D$ & 4.10 & 5.72 \\
$D\bar D^*+\mathrm{c.c.}$ & 0.95 & 0.22 \\
$D_s\bar D_s$ & 15.20 & 5.48 \\
$D_s\bar D_s^*+\mathrm{c.c.}$ & 68.61 & 33.82 \\
\end{tabular}
\end{ruledtabular}
\end{table}

\subsection{Threshold-reduced flavor--spin non-factorization}
\label{sec:nonfactorization}

To test whether the spin replacement $PP\to PV$ and the flavor replacement
$nonstrange\to strange$ act independently after removing the common
threshold factor, we construct the reduced response matrix
\begin{equation}
\widehat W=
\begin{pmatrix}
\widehat\Gamma_{n,PP}&\widehat\Gamma_{n,PV}\\
\widehat\Gamma_{s,PP}&\widehat\Gamma_{s,PV}
\end{pmatrix},
\end{equation}
where
\begin{equation}
\widehat\Gamma_{f,c}
=
\frac{\Gamma_{f,c}}
{\sum_{a\in(f,c)}|\bm P_{f,a}|^3}.
\end{equation}
If the flavor and spin dependences factorized as
$\widehat\Gamma_{f,c}\simeq F_fS_c$, this matrix would have rank one.
We therefore define the bounded determinant-based quantity
\begin{equation}
\mathcal N_{FS}
=
\frac{
\widehat\Gamma_{n,PP}\widehat\Gamma_{s,PV}
-
\widehat\Gamma_{n,PV}\widehat\Gamma_{s,PP}
}{
\widehat\Gamma_{n,PP}\widehat\Gamma_{s,PV}
+
\widehat\Gamma_{n,PV}\widehat\Gamma_{s,PP}
},
\label{eq:flavor_spin_nonfactorization}
\end{equation}
with $-1\leq\mathcal N_{FS}\leq1$.
It vanishes for the factorization null hypothesis and is invariant under
independent overall normalizations of the strange and nonstrange rows.
This quantity is a threshold-reduced structural measure and is not meant
to replace the experimentally oriented observable $A_{sn}$.

\begin{figure}[htbp]
\centering
\includegraphics[width=0.72\textwidth]{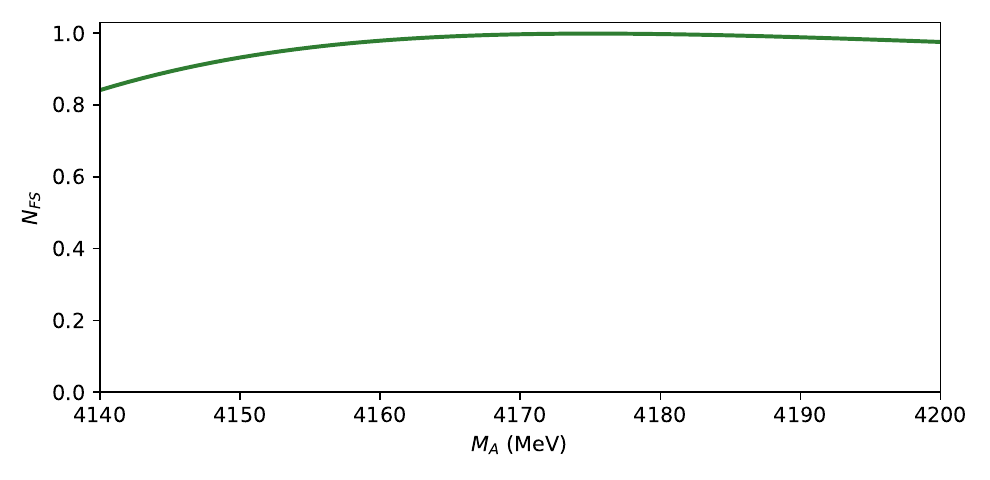}
\caption{Threshold-stripped flavor--spin non-factorization diagnostic $\mathcal N_{FS}$ of Eq.~\eqref{eq:flavor_spin_nonfactorization} over the fixed-wave-function mass scan. The horizontal line is the factorization null value.}
\label{fig:flavor_spin_nonfactorization}
\end{figure}

At $4146~\MeV$, the unrounded reduced strengths give
\[
(\widehat\Gamma_{n,PP},
\widehat\Gamma_{n,PV},
\widehat\Gamma_{s,PP},
\widehat\Gamma_{s,PV})
=
(4.098,0.949,15.204,68.605),
\]
in common normalized units, yielding
\[
\mathcal N_{FS}=0.902.
\]
Figure~\ref{fig:flavor_spin_nonfactorization} shows that
$\mathcal N_{FS}$ remains strongly positive throughout the fixed-wave-function
mass scan, ranging from $0.842$ at $4140~\MeV$ to $0.999$ near the
nonstrange $PV$ minimum. Thus, after removing the common $P$-wave
threshold behavior, the spin replacement $PP\to PV$ exhibits a strongly
flavor-dependent dynamical response in the pure-$2D$ realization.

\begin{table}[t]
\caption{Auxiliary cross-flavor and double-ratio diagnostics retained for comparison with the purely kinematic $P$-wave reference. These quantities are not used as the primary experimental discriminator in the main text.}
\label{tab:auxiliary_ratios}
\begin{ruledtabular}
\begin{tabular}{lcccc}
Quantity & Full, $4146$ & Threshold ref., $4146$ & Full, $4191$ & Threshold ref., $4191$ \\
\hline
$R_{PP}^{s/n}$ & 0.70 & 0.19 & 0.21 & 0.22 \\
$R_{PV}^{s/n}$ & 4.53 & 0.06 & 16.73 & 0.11 \\
$\mathcal D=R_s/R_n$ & 6.44 & 0.33 & 80.46 & 0.50 \\
\end{tabular}
\end{ruledtabular}
\end{table}

\begin{table}[t]
\caption{Strange-pair normalization check. The baseline prescription is $g_s=g_n$; the alternative prescription keeps the conventional dimensionless strength flavor independent, $\gamma_s=\gamma_n$. Entries marked ``same'' are unchanged by the overall strange-pair normalization.}
\label{tab:strange_prescription}
\begin{ruledtabular}
\begin{tabular}{lccc}
Observable & $g_s=g_n$ & $\gamma_s=\gamma_n$ & Sensitive? \\
\hline
$\Gamma(D_s\bar D_s)$ & 4.20 & 11.07 & yes \\
$\Gamma(D_s\bar D_s^*+\mathrm{c.c.})$ & 6.85 & 18.05 & yes \\
$R_s$ & 1.6298 & same & no \\
$R_{PP}^{s/n}$ & 0.7035 & 1.8541 & yes \\
$R_{PV}^{s/n}$ & 4.5337 & 11.9480 & yes \\
$C_{D_s\bar D_s}$ & $0.4082$ & same & no \\
$C_{D_s\bar D_s^*}$ & $0.1554$ & same & no \\
\end{tabular}
\end{ruledtabular}
\end{table}

\begin{figure}[htbp]
\centering
\includegraphics[width=0.7\textwidth]{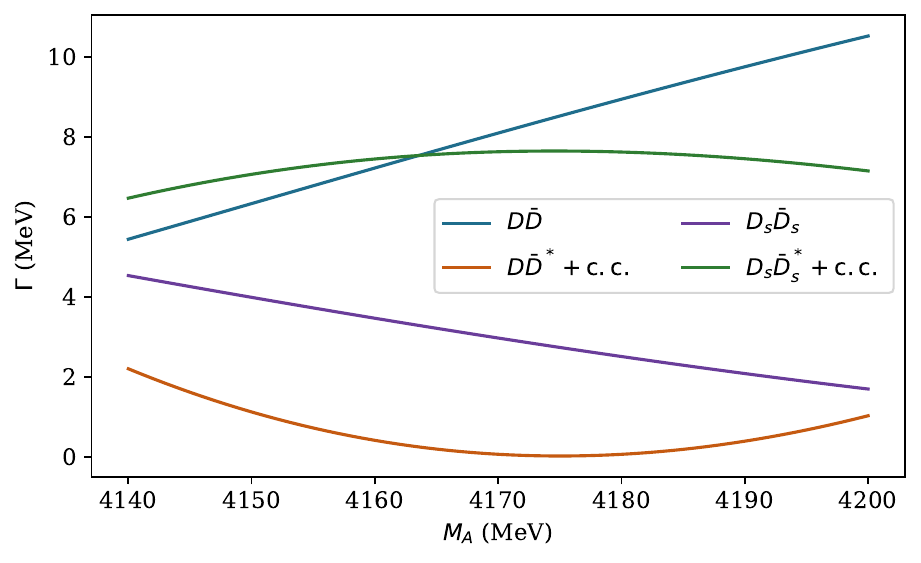}
\caption{Charge-summed widths vs external mass. The fixed-wave-function baseline gives a sampled $D\bar D^*+\mathrm{c.c.}$ minimum of about $0.02367~\MeV$ at $4175~\MeV$. Representative neutral and charged sectors have separate near-zeros near $4171$ and $4179~\MeV$ and are added incoherently; the charge-summed minimum is not a cancellation between charge states.}
\label{fig:phase_space_comparison}
\end{figure}

For reference, the charge-summed widths over the external-mass scan are shown in Fig.~\ref{fig:phase_space_comparison}.

\begin{table}[t]
\caption{Numerical comparison for $\psi(4160)\equiv2\,{}^3D_1$. BGS abbreviates $D\bar D^*+\bar D D^*$ as $DD^*$ and uses charge-averaged masses; Gui \emph{et al.} list neutral and charged sectors separately, which are summed here without an additional c.c. factor. The Gui table labels the state as $\psi(4160)$ but does not print an initial mass.}
\label{tab:model_comparison}
\begin{ruledtabular}
\begin{tabular}{lccccc}
Model & $M_A$ (MeV) & $\Gamma(DD)$ & $\Gamma(DD^*\mathrm{+c.c.})$ & $R_n$ & Convention \\
\hline
BGS~\cite{Barnes2005} & 4159 & $16.0$ & $0.4$ & $0.025$ & charge averaged \\
Gui LP~\cite{Gui2018} & $\psi(4160)$ & $12.0$ & $2.6$ & $0.217$ & charge summed \\
Gui SP~\cite{Gui2018} & $\psi(4160)$ & $13.9$ & $6.6$ & $0.475$ & charge summed \\
This work & 4146 & $5.97$ & $1.51$ & $0.253$ & charge summed \\
This work & 4191 & $9.83$ & $0.44$ & $0.0452$ & charge summed \\
\end{tabular}
\end{ruledtabular}
\end{table}

\FloatBarrier

\section{Extended mixing and wave-function diagnostics}
\label{app:stability}
The main text gives the real two-state amplitude and its channel-resolved bilinears. The extended scans below quantify sensitivity to an additional relative phase and to wave-function shape.

\subsection{Coherent admixture and phase sensitivity}

For completeness, the phase dependence beyond the real two-state ansatz is made explicit by introducing $\phi$ between the $3S$ and $2D$ basis amplitudes:
\begin{equation}
|\psi\rangle=\cos\theta\,|2D\rangle+e^{i\phi}\sin\theta\,|3S\rangle.
\label{eq:complex_mixing}
\end{equation}
Using the bilinears defined in Eq.~\eqref{eq:interference_definitions}, the corresponding width is
\begin{equation}
\Gamma_i(\theta,\phi)=D_i\cos^2\theta+S_i\sin^2\theta
+2X_i\sin\theta\cos\theta\cos\phi.
\label{eq:complex_mixing_width}
\end{equation}
It is an algebraic phase-sensitivity exercise; the relative phase is not determined in the present calculation. The $\phi=0$ and $\phi=\pi$ slices are the two real-branch limits, while the continuous scan provides a sensitivity envelope rather than a selected physical phase.

\begin{figure}[htbp]
\centering
\includegraphics[width=0.88\textwidth]{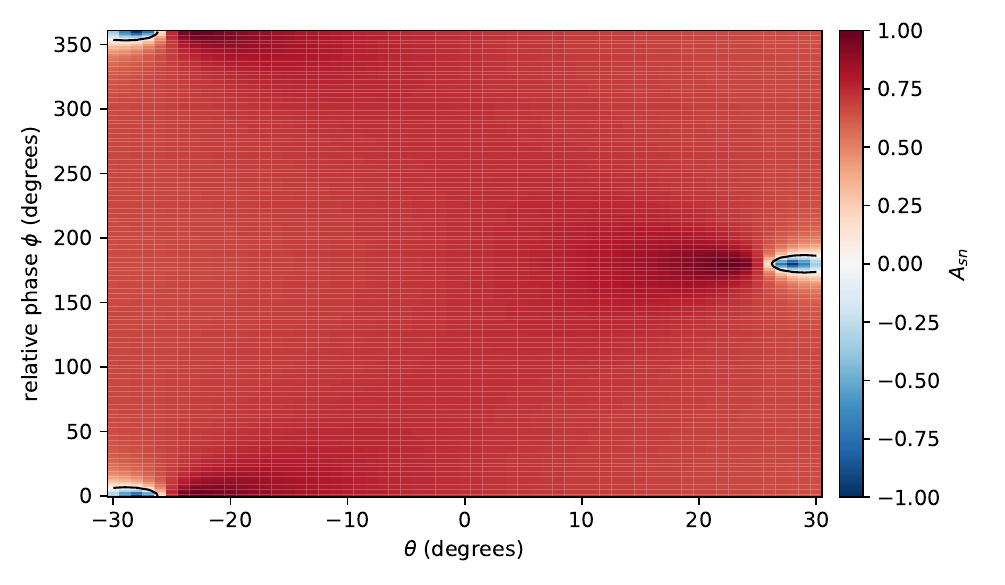}
\caption{Phase and coherent-admixture dependence of $A_{sn}$ at $M_A=4146~\MeV$, obtained from the bilinears in Table~\ref{tab:interference_bilinears}.}
\label{fig:mixing_phase_map}
\end{figure}

Figure~\ref{fig:mixing_phase_map} displays the resulting algebraic sensitivity of $A_{sn}$ to the two-component ansatz. The negative region is confined to narrow destructive-interference islands, where one or both same-flavor ratios are small, reinforcing its interpretation as an interference-sensitive feature.

\begin{table}[t]
\caption{Charge-summed partial widths with coherent $3S$--$2D$ mixing at $M_A=4146~\MeV$. The scan uses the relative signs of the supplied $2D$ and $3S$ solutions; the sign of $\theta$ changes convention if either basis vector is rephased. Displayed widths are rounded; derived ratios use the unrounded output.}
\label{tab:widths_mixing}
\begin{ruledtabular}
\begin{tabular}{lccccccc}
Channel & $\Gamma_i(-30^\circ)$ & $\Gamma_i(-20^\circ)$ & $\Gamma_i(-10^\circ)$ & $\Gamma_i(0^\circ)$ & $\Gamma_i(10^\circ)$ & $\Gamma_i(20^\circ)$ & $\Gamma_i(30^\circ)$ \\
\hline
$D\bar D$ & 10.99 & 9.71 & 7.97 & 5.97 & 3.96 & 2.18 & 0.84 \\
$D\bar D^*+\mathrm{c.c.}$ & 0.15 & 0.03 & 0.50 & 1.51 & 2.94 & 4.61 & 6.32 \\
$D_s\bar D_s$ & 7.56 & 6.72 & 5.56 & 4.20 & 2.82 & 1.59 & 0.64 \\
$D_s\bar D_s^*+\mathrm{c.c.}$ & 0.05 & 0.57 & 2.93 & 6.85 & 11.84 & 17.31 & 22.60 \\
\end{tabular}
\end{ruledtabular}
\end{table}

\subsection{Momentum-space and wave-function-shape diagnostics}

The cumulative amplitudes complement the local integrands shown in the main text:

\begin{figure}[htbp]
\centering
\includegraphics[width=0.85\textwidth]{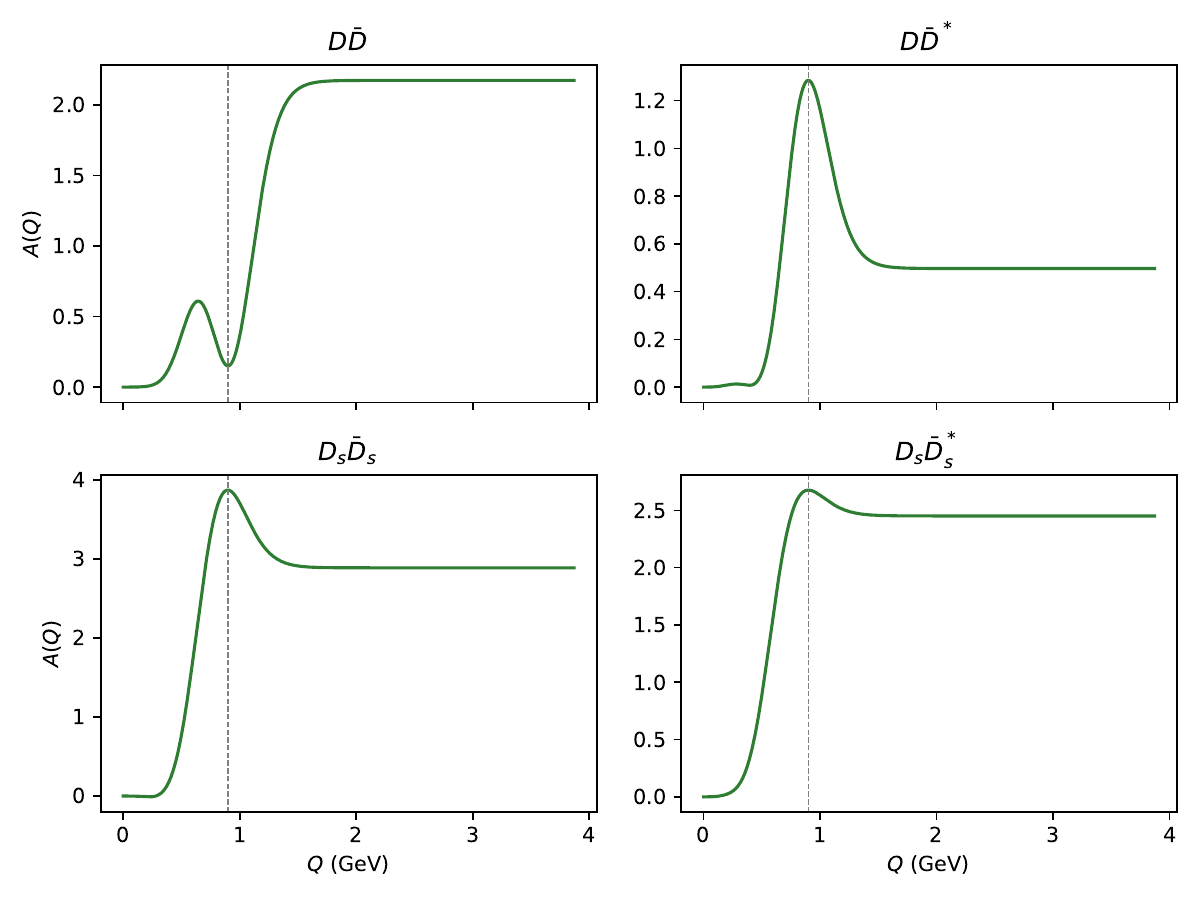}
\caption{Four-channel cumulative overlap amplitudes at $M_A=4146~\MeV$ in the same $2\times2$ arrangement as Fig.~\ref{fig:integrands}. A strong decrease after the nodal region signals destructive cancellation in the full radial integral.}
\label{fig:cumulative}
\end{figure}

\begin{figure}[htbp]
\centering
\includegraphics[width=0.85\textwidth]{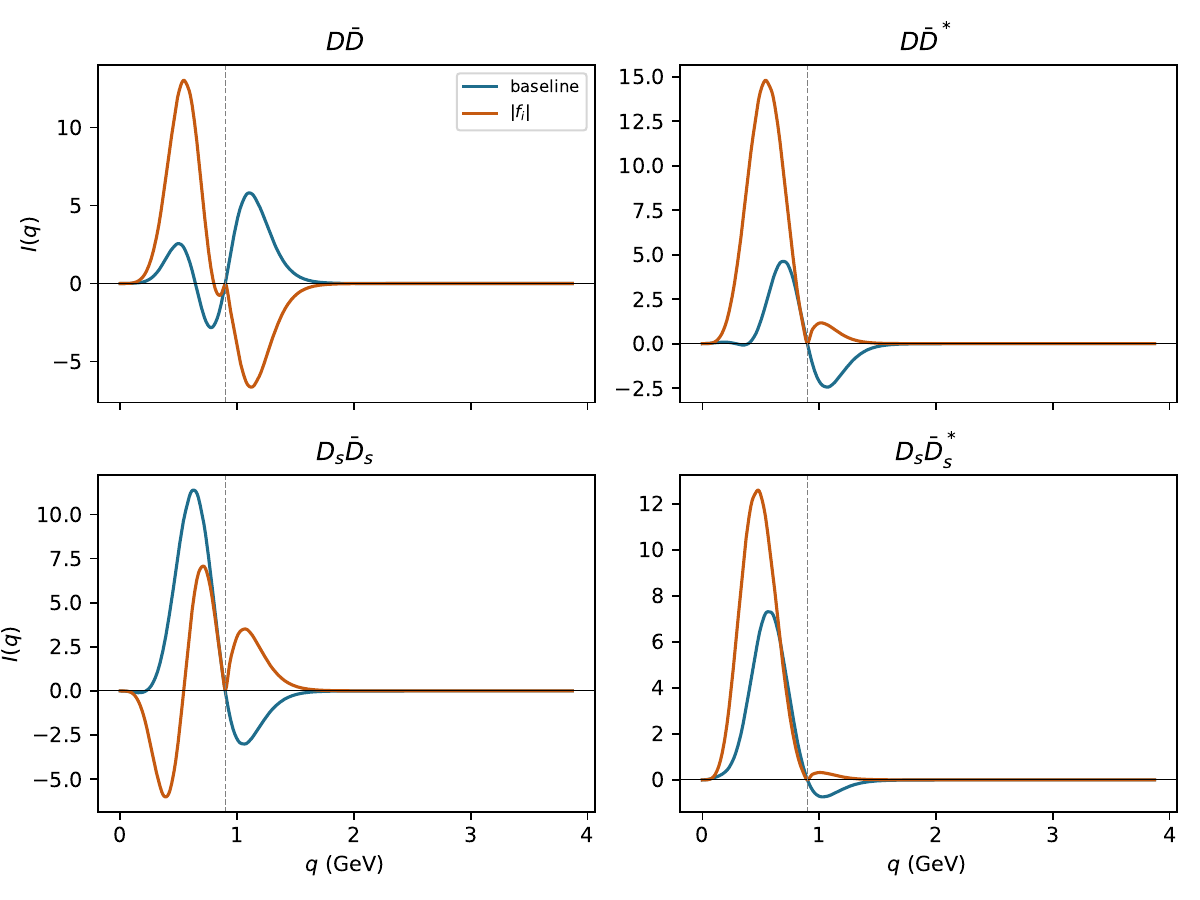}
\caption{Comparison between the baseline $2D$ integrands and the diagnostic component-level $f_i\to|f_i|$ deformation at $M_A=4146~\MeV$. Under this prescription the $PV$ integrands are numerically sign definite within integration precision and their endpoint amplitudes increase, whereas the $PP$ endpoint amplitudes decrease and their normalized cancellation indices increase.}
\label{fig:node_removed}
\end{figure}

The component-level sign-removal test quoted in the main text is shown explicitly in Fig.~\ref{fig:node_removed}.

\begin{figure}[htbp]
\centering
\includegraphics[width=0.86\textwidth]{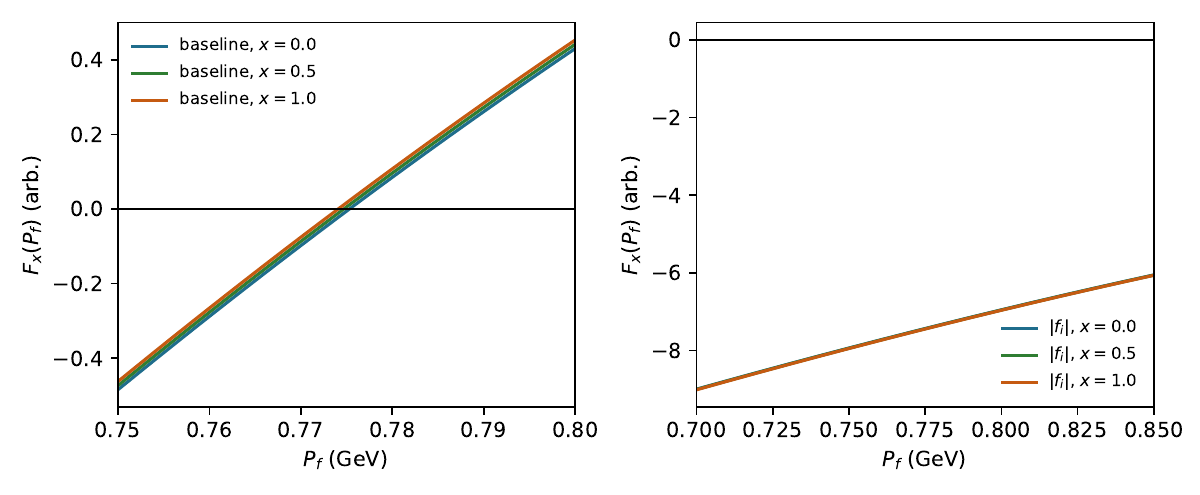}
\caption{Threshold-reduced projected amplitudes for representative continuation paths. Left: baseline signed $2D$ input, with zeros in the physical recoil window. Right: the diagnostic $f_\alpha\to|f_\alpha|$ counterfactual over the wider recoil interval. The latter has no sign change in the displayed $0.70$--$0.85~\GeV$ range. The counterfactual is not a self-consistent Salpeter solution.}
\label{fig:recoil_sign_removal}
\end{figure}

As a further diagnostic, the recoil-zero structure can be tested for its
dependence on the signed input. We repeat the charged--neutral
continuation with the existing component-level counterfactual
$f_\alpha(q)\to|f_\alpha(q)|$, reconstructing the dependent Salpeter
components with the same production prescription. This is not a
self-consistent eigenstate and is not used as an uncertainty estimate.
It is a sign-pattern test. The counterfactual is scanned over the wider
interval $0.70\leq P_f\leq0.85~\GeV$ for all eleven continuation paths.
No projected-amplitude sign change is found in this interval, whereas
every baseline path has a zero in the narrower physical recoil window.
Figure~\ref{fig:recoil_sign_removal} shows representative paths. The
absence of the common recoil zero in this counterfactual supports the
statement that the recoil-space cancellation in the present realization
depends on the signed multi-component radial structure. It does not
prove that a radial node is the unique cause of the physical decay
pattern.

\begin{figure}[htbp]
\centering
\includegraphics[width=0.90\textwidth]{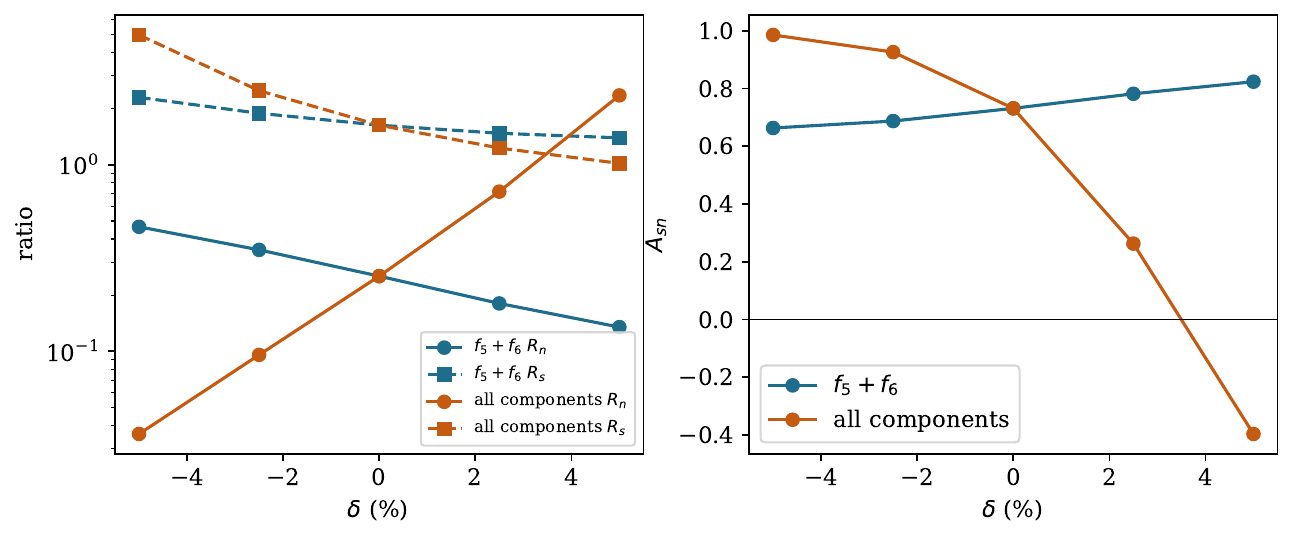}
\caption{Diagnostic sensitivity to node-position deformations at $M_A=4146~\MeV$. Left: $R_n$ and $R_s$ for the restricted $f_5+f_6$ deformation and the all-component deformation. Right: the corresponding bounded asymmetry $A_{sn}$. The restricted prescription preserves the baseline ordering over $\pm5\%$, whereas the all-component prescription substantially weakens it at $+2.5\%$ ($A_{sn}=0.263$, $R_s/R_n=1.71$) and reverses it at $+5\%$ ($A_{sn}=-0.396$).}
\label{fig:node_stability}
\end{figure}

To test the sensitivity to the internal node position, we construct a multiplicative diagnostic deformation of the sign-changing Salpeter components. A zero-point audit using fixed-window integral weights and local peak ratios (Table~\ref{tab:component_zero_audit}) identifies dominant resolved sign changes in $f_5$ and $f_6$ and subleading tail sign changes in $f_3$ and $f_4$ with substantially smaller post-zero support. For each independent component $f_i(q)$ with a numerical zero at $q_{0,i}$, we define the fixed-window ($\Delta=0.20~\GeV$) local peak ratios
\begin{equation}
\rho_i^- = \frac{\max_{q\in[q_{0,i}-\Delta,q_{0,i})}|f_i(q)|}{\max_q|f_i(q)|},
\qquad
\rho_i^+ = \frac{\max_{q\in(q_{0,i},q_{0,i}+\Delta]}|f_i(q)|}{\max_q|f_i(q)|},
\end{equation}
and the fixed-window integral weights
\begin{equation}
W_i^- = \frac{\int_{q_{0,i}-\Delta}^{q_{0,i}} dq\,q^2|f_i(q)|^2}{\int_0^{q_{\max}} dq\,q^2|f_i(q)|^2},
\qquad
W_i^+ = \frac{\int_{q_{0,i}}^{q_{0,i}+\Delta} dq\,q^2|f_i(q)|^2}{\int_0^{q_{\max}} dq\,q^2|f_i(q)|^2}.
\end{equation}
A sign change is classified as resolved when both sides of the zero carry non-negligible local amplitude or integrated weight within the fixed window. The numerical grid extends to $q_{\max}=3.87~\GeV$. The sign changes of $f_3$ and $f_4$ have substantially smaller post-zero support than those of $f_5$ and $f_6$. We therefore use the restricted $f_5+f_6$ deformation and the all-component deformation as two diagnostic prescriptions. Neither has a privileged statistical status. Let $\mathcal S$ denote the set of selected independent components. For each $f_j$ with a numerical zero at $q_{0,j}$, let $\delta$ denote the fractional shift of that zero. The shifted component is defined as
\[
f_j^{(\delta)}(q)=\mathcal N_j\,f_j(q)\,\frac{q-q_{0,j}'}{q-q_{0,j}},
\qquad q_{0,j}'=q_{0,j}(1+\delta),
\]
where the normalization factor is
\[
\mathcal N_j=\left[
\frac{\int_0^{q_{\max}}dq\,q^2|f_j(q)|^2}
{\int_0^{q_{\max}}dq\,q^2\left|f_j(q)(q-q_{0,j}')/(q-q_{0,j})\right|^2}
\right]^{1/2}>0.
\]
For the restricted prescription, $\mathcal S_{\rm res}=\{5,6\}$; for the all-component prescription, $\mathcal S_{\rm all}=\{3,4,5,6\}$. Components outside $\mathcal S$ are unchanged. The dependent Salpeter structures are reconstructed from the deformed independent functions using the same standard Salpeter constraint relations~\cite{Chang2005,Chang2004,Wang2006}, with cubic interpolation on the radial grid. At $\delta=0$, both prescriptions reduce to the undeformed baseline. At the original zero, the shifted component is evaluated from the finite analytic limit
\[
f_j^{(\delta)}(q_{0,j})=\mathcal N_j\,f_j'(q_{0,j})\left(q_{0,j}-q_{0,j}'\right),\qquad j\in\mathcal S,
\]
rather than by evaluating the singular ratio separately. This is a controlled shape-sensitivity test rather than a self-consistent new Salpeter solution. The scan uses the five deformation points shown in Table~\ref{tab:node_scan} and Fig.~\ref{fig:node_stability}.

\begin{table}[t]
\caption{Zero-point audit of the four independent Salpeter components, using fixed-window integral weights ($W^\pm$) and local peak ratios ($\rho^\pm$) at $\Delta=0.20~\GeV$. Both $f_5$ and $f_6$ have dominant resolved sign changes; $f_3$ and $f_4$ have subleading tail sign changes with substantially smaller post-zero support. The distinction is quantitative rather than binary.}
\label{tab:component_zero_audit}
\begin{ruledtabular}
\begin{tabular}{lcccccc}
Component & $q_{0,i}$ (GeV) & $\rho^-_{0.20}$ & $\rho^+_{0.20}$ & $W^-_{0.20}$ & $W^+_{0.20}$ & Classification \\
\hline
$f_3$ & $0.899$ & $1.00\times10^{-1}$ & $2.44\times10^{-2}$ & $1.35\times10^{-2}$ & $3.04\times10^{-3}$ & subleading tail \\
$f_4$ & $0.898$ & $1.02\times10^{-1}$ & $2.55\times10^{-2}$ & $1.37\times10^{-2}$ & $3.28\times10^{-3}$ & subleading tail \\
$f_5$ & $0.900$ & $5.67\times10^{-1}$ & $3.21\times10^{-1}$ & $5.54\times10^{-2}$ & $4.24\times10^{-2}$ & dominant resolved \\
$f_6$ & $0.903$ & $5.54\times10^{-1}$ & $2.83\times10^{-1}$ & $6.46\times10^{-2}$ & $4.06\times10^{-2}$ & dominant resolved \\
\end{tabular}
\end{ruledtabular}
\end{table}

For compactness in the following appendix tables, we use $\Gamma_{DD}\equiv\Gamma(D\bar D)$, $\Gamma_{DD^*}\equiv\Gamma(D\bar D^*+\mathrm{c.c.})$, $\Gamma_{D_sD_s}\equiv\Gamma(D_s\bar D_s)$, and $\Gamma_{D_sD_s^*}\equiv\Gamma(D_s\bar D_s^*+\mathrm{c.c.})$.

\begin{table}[t]
\caption{Node-position deformation at $M_A=4146~\MeV$. The resolved sign-changing components $f_5$ and $f_6$ are jointly shifted; $f_3,f_4$ are held fixed. All widths are charge-summed, in MeV.}
\label{tab:node_scan}
\begin{ruledtabular}
\begin{tabular}{cccccccc}
$\delta$ (\%) & $\Gamma_{DD}$ & $\Gamma_{DD^*}$ & $\Gamma_{D_sD_s}$ & $\Gamma_{D_sD_s^*}$ & $R_n$ & $R_s$ & $A_{sn}$ \\
\hline
$-5$ & $9.75$ & $4.54$ & $3.27$ & $7.52$ & $0.465$ & $2.297$ & $0.663$ \\
$-2.5$ & $7.62$ & $2.67$ & $3.77$ & $7.12$ & $0.350$ & $1.888$ & $0.687$ \\
$0$ & $5.97$ & $1.51$ & $4.20$ & $6.85$ & $0.253$ & $1.630$ & $0.731$ \\
$2.5$ & $4.75$ & $0.86$ & $4.54$ & $6.70$ & $0.181$ & $1.475$ & $0.782$ \\
$5$ & $3.88$ & $0.52$ & $4.78$ & $6.66$ & $0.135$ & $1.394$ & $0.824$ \\
\end{tabular}
\end{ruledtabular}
\end{table}

\begin{table}[t]
\caption{Comparison of restricted resolved-component and all-component diagnostic node deformations at $M_A=4146~\MeV$. All widths are charge-summed, in MeV. The restricted prescription preserves the ordering throughout $\pm5\%$, while the all-component prescription substantially weakens it at $\delta=+2.5\%$ ($A_{sn}=0.263$, $R_s/R_n=1.71$) and reverses it at $\delta=+5\%$ ($A_{sn}=-0.396$).}
\label{tab:component_selection}
\begin{ruledtabular}
\begin{tabular}{cccccccc}
Scenario & $\delta$ (\%) & $\Gamma_{DD}$ & $\Gamma_{DD^*}$ & $\Gamma_{D_sD_s}$ & $\Gamma_{D_sD_s^*}$ & $A_{sn}$ & Hierarchy \\
\hline
$f_5+f_6$ & $-5$ & $9.75$ & $4.54$ & $3.27$ & $7.52$ & $0.663$ & $R_n<R_s$ \\
$f_5+f_6$ & $-2.5$ & $7.62$ & $2.67$ & $3.77$ & $7.12$ & $0.687$ & $R_n<R_s$ \\
$f_5+f_6$ & $0$ & $5.97$ & $1.51$ & $4.20$ & $6.85$ & $0.731$ & $R_n<R_s$ \\
$f_5+f_6$ & $+2.5$ & $4.75$ & $0.86$ & $4.54$ & $6.70$ & $0.782$ & $R_n<R_s$ \\
$f_5+f_6$ & $+5$ & $3.88$ & $0.52$ & $4.78$ & $6.66$ & $0.824$ & $R_n<R_s$ \\
all four & $-5$ & $16.41$ & $0.59$ & $1.21$ & $6.02$ & $0.986$ & $R_n<R_s$ \\
all four & $-2.5$ & $10.29$ & $0.98$ & $2.57$ & $6.42$ & $0.926$ & $R_n<R_s$ \\
all four & $0$ & $5.97$ & $1.51$ & $4.20$ & $6.85$ & $0.731$ & $R_n<R_s$ \\
all four & $+2.5$ & $3.12$ & $2.24$ & $5.95$ & $7.32$ & $0.263$ & $R_n<R_s$ \\
all four & $+5$ & $1.38$ & $3.25$ & $7.70$ & $7.84$ & $-0.396$ & $R_n>R_s$ \\
\end{tabular}
\end{ruledtabular}
\end{table}

The two node-shift prescriptions give the same undeformed baseline but diverge for sufficiently large positive shifts. The restricted $f_5+f_6$ prescription preserves $R_n<R_s$ over $\pm5\%$, whereas the all-component prescription weakens the ordering at $\delta=+2.5\%$ and reverses it at $\delta=+5\%$. This stress test shows that the matched ratios probe the detailed distribution of the relativistic multi-component wave function around the sign-changing region, rather than the mere existence of a node.

The full counterfactual cancellation results are summarized below.

\begin{table}[t]
\caption{Diagnostic comparison of the baseline $2D$ solution with the component-level $f_i\to|f_i|$ sign-pattern deformation at $M_A=4146$ and $4191~\MeV$. The deformation preserves the norms of the independent components but changes relative signs, phases, and reconstructed structures. For $PV$ channels the representative $M_{12}$ component is reported. The superscripts $(2D)$ and $(|f|)$ denote the representative projected amplitudes before and after the sign-removal deformation, respectively. The index $C_i$ and response $L_i=|M_i^{(|f|)}|/|M_i^{(2D)}|$ characterize that projected amplitude rather than the charge-summed channel.}
\label{tab:node_removed}
\begin{ruledtabular}
\begin{tabular}{lcccccc}
& \multicolumn{3}{c}{$M_A=4146~\MeV$} & \multicolumn{3}{c}{$M_A=4191~\MeV$} \\
\cline{2-4}\cline{5-7}
Channel & $C_i^{(2D)}$ & $C_i^{(|f|)}$ & $L_i$ & $C_i^{(2D)}$ & $C_i^{(|f|)}$ & $L_i$ \\
\hline
$D\bar D$ & 0.2970 & 0.7334 & 0.79 & 0.1330 & 0.7100 & 0.72 \\
$D\bar D^*+\mathrm{c.c.}$ & 0.7617 & 0.0000 & 12.07 & 0.7926 & 0.0000 & 15.11 \\
$D_s\bar D_s$ & 0.4082 & 0.6872 & 0.45 & 0.5420 & 0.9082 & 0.20 \\
$D_s\bar D_s^*+\mathrm{c.c.}$ & 0.1554 & 0.0000 & 1.99 & 0.2261 & 0.0000 & 2.47 \\
\end{tabular}
\end{ruledtabular}
\end{table}

\FloatBarrier

\section{Detailed limitations and numerical checks}
\label{app:limits}
The spectroscopic input supplies the nearby $3S$ and $2D$ BS basis states used throughout this work. The difference between the calculated bare masses and the physical-region external masses sets the scale for the effective two-state organization. The decay calculation also inherits the usual model dependence of the pair-creation strength, wave-function normalization, and resonance mass input.

The external-mass scan isolates sensitivity to threshold factors, recoil momentum, and the mass-dependent positive-energy coefficient reconstruction. The tabulated radial functions remain fixed, so the scan does not include the change of those functions that would arise from retuning the interaction kernel to reproduce a different bound-state mass.

The relative normalization of strange and nonstrange absolute widths is model dependent because $g_s$ is not independently calibrated by the $\psi(3770)$ decay. This uncertainty affects absolute strange-channel widths and cross-flavor ratios, but it does not affect the same-flavor ratios, the bounded asymmetry $A_{sn}$, or the cancellation indices. The Salpeter constraints reconstruct the positive-energy amplitudes from the independent radial functions, and all reported results use the instantaneous trace specified in Sec.~\ref{sec:formalism}. The zero locations should be interpreted as features of this specified projection rather than precision mass predictions. On the central grid, the representative physical neutral and charged zeros occur at approximately $4171.3$ and $4178.7~\MeV$, with recoil momenta $0.77530$ and $0.77537~\GeV$, respectively. Across the coarse, central, and fine grids, the zero masses vary by no more than $0.12~\MeV$ and the recoil momenta by no more than $0.16~\MeV$; these figures quantify numerical interpolation stability within the stated projection, not the full theoretical uncertainty.

The real-angle scan is reported in the main text; the additional relative-phase and wave-function-deformation scans are reported in Appendix~\ref{app:stability}. Here we focus on the implementation limits and numerical convergence of the pure-$2D$ component calculation.

\subsection{Kinematic charged--neutral interpolation}
\label{app:kinematic_interpolation}

\begin{figure}[htbp]
\centering
\includegraphics[width=0.88\textwidth]{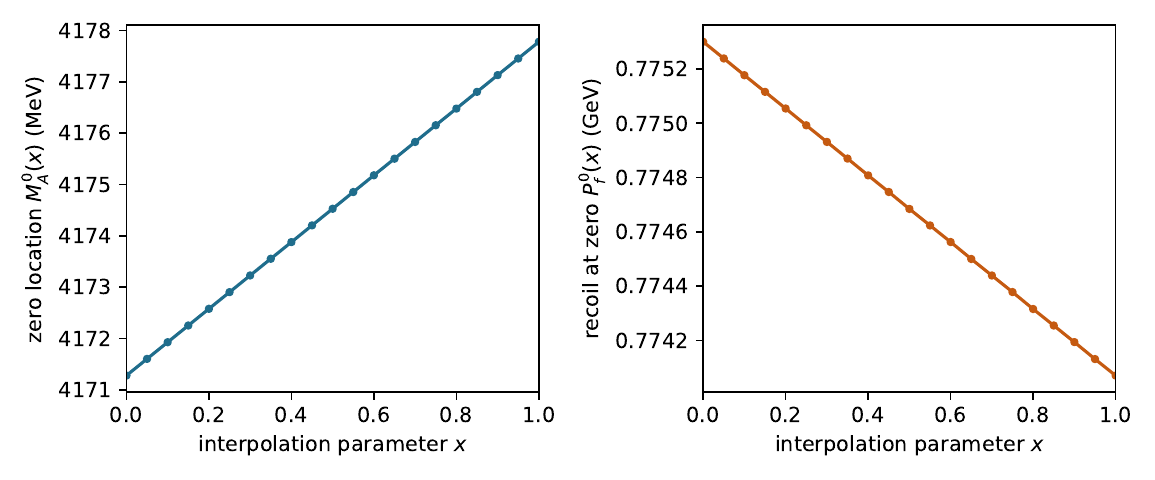}
\caption{Continuous external-mass interpolation from neutral to charged $D$ and $D^*$ masses with fixed neutral-sector radial inputs. Left: external mass at the projected-amplitude zero. Right: recoil momentum at that zero. The continuation is a kinematic test, not a family of physical mesons or self-consistent bound states.}
\label{fig:charged_neutral_interpolation}
\end{figure}

The kinematic continuation introduced in Sec.~\ref{sec:recoil_zeros} interpolates the external $D$ and $D^*$ masses while keeping the neutral-sector radial inputs and all model parameters fixed. Intermediate values of $x$ therefore do not represent physical mesons or self-consistent bound-state solutions. The external-mass zero moves from $4171.28$ to $4177.78~\MeV$, whereas its recoil momentum changes only from $0.77530$ to $0.77407~\GeV$. The full trajectory is shown in Fig.~\ref{fig:charged_neutral_interpolation}.

\subsection{Self-consistent heavy-light continuation}
\label{app:self_consistent_recoil}

The kinematic continuation leaves the final-state radial functions fixed. We therefore perform a separate self-consistent strong-model continuation, $m_q(x)=0.305+0.006x~\GeV$, with $x=0,0.2,\ldots,1$. At each point the pseudoscalar and vector heavy-light Salpeter equations are solved independently using the same kernel prescription, and the resulting eigenmasses and radial components are inserted into the unchanged decay projector. This is not an electromagnetic interpolation of physical mesons; it tests the recoil-zero condition along a continuous family of bound-state solutions of the present model.

\begin{figure}[t]
\centering
\includegraphics[width=0.88\textwidth]{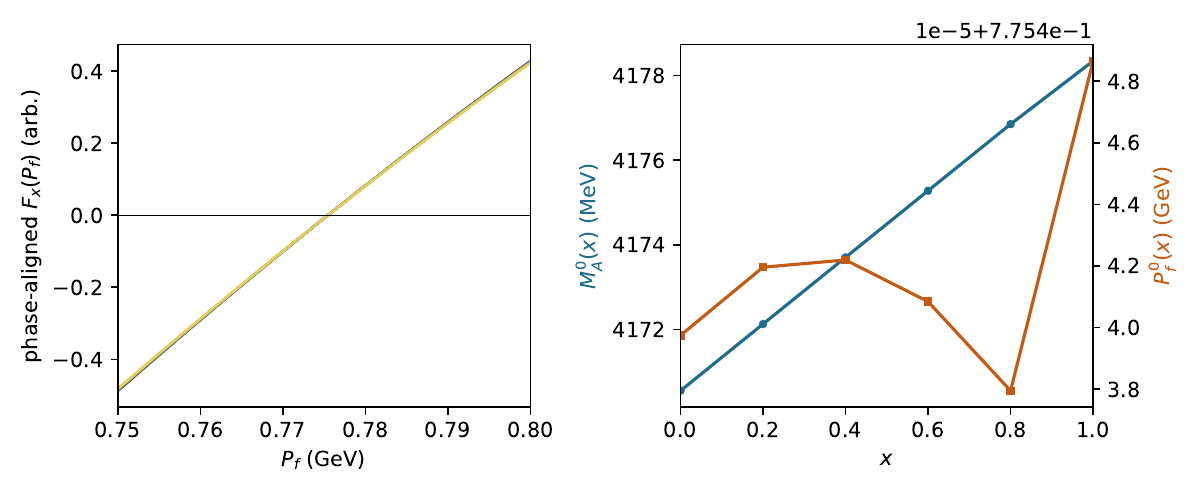}
\caption{Self-consistent heavy-light continuation, with $m_q(x)=0.305+0.006x~\GeV$. Left: phase-aligned threshold-reduced projected amplitudes. Right: mass and recoil coordinates of their zeros; the blue-circle curve gives $M_A^0(x)$ on the left axis, while the orange-square curve gives $P_f^0(x)$ on the right axis. Global signs are aligned only to compare the shapes of independently solved eigenvectors and do not alter the zeros.}
\label{fig:self_consistent_recoil}
\end{figure}

Figure~\ref{fig:self_consistent_recoil} shows that the zero mass changes from $4170.55$ to $4178.34~\MeV$, while the recoil zero remains within a spread $\Delta P_f^0=1.07\times10^{-5}~\GeV$, or $0.00138\%$ of its mean value. Since independently solved eigenvectors carry arbitrary global signs, the threshold-reduced curves are phase-aligned before comparing their shapes. Their local collapse metric is $0.00231$. The phase alignment does not affect any zero location. Thus the stronger continuation test preserves the recoil organization of the cancellation within the current positive-energy instantaneous Salpeter realization.

\subsection{Expanded radial-support check}
\label{app:radial_support}

\begin{figure}[t]
\centering
\includegraphics[width=0.88\textwidth]{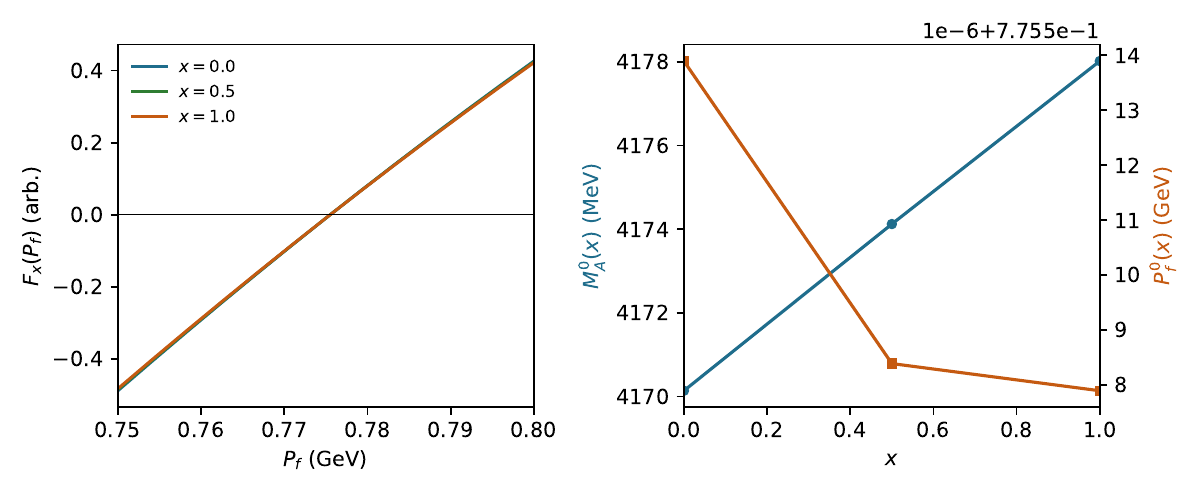}
\caption{Expanded-support check using independently solved final-state wave functions with $N_1=257$ and $q_{\max}=7.71~\GeV$. Left: phase-aligned threshold-reduced projected amplitudes. Right: zero locations along the self-consistent light-mass trajectory; the blue-circle curve gives $M_A^0(x)$ on the left axis, while the orange-square curve gives $P_f^0(x)$ on the right axis.}
\label{fig:radial_support}
\end{figure}

The preceding continuation uses final-state radial grids with $129$ points and $q_{\max}=3.87~\GeV$. To test sensitivity to their ultraviolet support, we re-solve the $D(x)$ and $D^*(x)$ equations at $x=0,0.5,1$ on $257$ points at the same spacing, extending the final-state support to $q_{\max}=7.71~\GeV$. The initial $\psi(4160)$ input and the decay projector are unchanged. Figure~\ref{fig:radial_support} gives a recoil-zero spread $5.996\times10^{-6}~\GeV$, or $0.000773\%$, and a phase-aligned local collapse metric $0.00277$. At the midpoint, enlarging the support shifts the zero from $P_f^0=0.775434$ to $0.775508~\GeV$. This small shift is well below the local recoil window and does not alter the recoil-controlled interpretation within the present model.

\subsection{Numerical convergence and input-grid support}
\label{app:numerical_convergence}

Table~\ref{tab:numerical_convergence} compares the benchmark calculation at three angular and radial integration grids. The $(72,56,8)$ grid is used for the central results. Within the fixed tabulated wave-function support, the four aggregate widths and the derived ratios are stable at the sub-percent level.

\begin{table}[t]
\caption{Numerical convergence at $M_A=4146~\MeV$ for the pure-$2D$ benchmark. Widths are charge summed and given in MeV; grid entries are $(n_q,n_{\cos\vartheta},n_\phi)$.}
\label{tab:numerical_convergence}
\begin{ruledtabular}
\begin{tabular}{lccccccc}
Grid & $\Gamma_{DD}$ & $\Gamma_{DD^*+\mathrm{c.c.}}$ & $\Gamma_{D_sD_s}$ & $\Gamma_{D_sD_s^*+\mathrm{c.c.}}$ & $R_n$ & $R_s$ & $A_{sn}$ \\
\hline
$(48,36,8)$ & 5.9721 & 1.4997 & 4.1901 & 6.8326 & 0.2511 & 1.6307 & 0.7331 \\
$(72,56,8)$ & 5.9731 & 1.5106 & 4.2023 & 6.8487 & 0.2529 & 1.6298 & 0.7313 \\
$(96,76,12)$ & 5.9482 & 1.5025 & 4.1919 & 6.8280 & 0.2526 & 1.6289 & 0.7315 \\
\end{tabular}
\end{ruledtabular}
\end{table}

The same three grids were also evaluated at the cancellation-sensitive masses $4171$, $4175$, and $4179~\MeV$. The charge-summed $D\bar D^*+\mathrm{c.c.}$ widths are stable at the percent level or better: respectively, $(0.0495,0.0510,0.0503)$, $(0.02374,0.02367,0.02356)$, and $(0.05227,0.05054,0.05085)$ MeV from coarse to fine grids. The individual neutral and charged near-zero sectors are correspondingly more sensitive in relative terms, as expected when the reference value is close to zero, but their small absolute scale and the charge-summed minimum persist. These checks establish quadrature stability of the reported charge-summed feature within the supplied radial input.

The tabulated radial functions decrease to zero at the available endpoint, $q_{\max}=3.87~\GeV$, and no extrapolation beyond that input support is introduced. The implementation therefore omits angular points for which a shifted final-state radial argument lies beyond the available grid. At the benchmark the skipped fractions range from about $6.6\%$ to $11.8\%$ across channels and remain stable under the grid changes in Table~\ref{tab:numerical_convergence}. Together with the vanishing endpoint trend of the supplied functions, this supports use of the finite grid for the present calculation; it is not an independent extrapolation test of a hypothetical high-momentum tail.

Remaining model uncertainties include the overall pair-creation normalization, the relative strange-pair prescription, potential and quark-mass parameters, and residual effects of the positive-energy decay projection. No full parameter-covariance propagation is attempted.

\subsection{Model limitations and scope}
\label{app:model_limitations}

Three boundaries are most relevant to cancellation-sensitive quantities:
the fixed-radial-function external-mass scan, the specified instantaneous
decay projection, and the finite radial support of the tabulated wave
functions. Their numerical implementation and convergence checks are
detailed in the preceding subsections. The self-consistent heavy-light
continuation is consistent with the recoil-zero organization within this
specified realization. Absolute strange-channel widths also depend on the
strange-pair prescription, whereas the primary same-flavor component
ratios and $A_{sn}$ do not. The all-component node deformation in
Appendix~\ref{app:stability} further shows that the baseline sign of
$A_{sn}$ is a diagnostic of the calculated BS shape, not an
arbitrary-deformation invariant.

The initial-state $2D$ and $3S$ radial functions are solved at their bare
masses of $4110$ and $4051~\MeV$ and reused at the external masses $4146$
and $4040~\MeV$, respectively. The fixed-wave-function scan therefore
tests the sensitivity of the overlap integrals to phase space and recoil
kinematics, but not to the change of the Salpeter components that would
accompany a retuned bound-state solution. For the $2D$ state, the
bare-to-external mass shift is $36~\MeV$, or about $0.9\%$ of the total
mass. The corresponding shape change of the multi-component wave function
is expected to be small relative to the $P$-wave threshold factor, but a
quantitative estimate is beyond the scope of the present calculation.


\begin{thebibliography}{99}

\bibitem{Barnes2005}
T. Barnes, S. Godfrey, and E. S. Swanson,
``Higher charmonia,''
Phys. Rev. D \textbf{72}, 054026 (2005) [arXiv:hep-ph/0505002].

\bibitem{GodfreyIsgur}
S. Godfrey and N. Isgur,
``Mesons in a relativized quark model with chromodynamics,''
Phys. Rev. D \textbf{32}, 189 (1985).

\bibitem{Brambilla}
N. Brambilla \textit{et al.} [Quarkonium Working Group],
``Heavy quarkonium physics,''
Rev. Mod. Phys. \textbf{77}, 1423 (2005) [arXiv:hep-ph/0412158].

\bibitem{Ke2026}
W.-Y. Ke, Q. Li, T.-H. Wang, T.-F. Feng, and G.-L. Wang,
``$S$-$P$-$D$ mixing in vector quarkonia from the Salpeter equation with optimized wave function representations,''
arXiv:2602.09692 [hep-ph].

\bibitem{Man2025vmm}
Z.~L.~Man, S.~Q.~Luo and X.~Liu,
``Is the 3S-2D mixing strong for the charmonia {\ensuremath{\psi}}(4040) and {\ensuremath{\psi}}(4160)?,''
Phys. Rev. D \textbf{112}, 074025 (2025)
[arXiv:2507.18536 [hep-ph]].

\bibitem{BESIII_DD_2024}
M. Ablikim \textit{et al.} [BESIII Collaboration],
``Precise measurement of Born cross sections for $e^+e^-\to D\bar D$ at $\sqrt{s}=3.80$--$4.95~\GeV$,''
Phys. Rev. Lett. \textbf{133}, 081901 (2024) [arXiv:2402.03829 [hep-ex]].

\bibitem{BESIII_DstarD_2022}
M. Ablikim \textit{et al.} [BESIII Collaboration],
``Cross section measurements of the $e^+e^-\to D^{*+}D^{*-}$ and $e^+e^-\to D^{*+}D^-$ processes at center-of-mass energies from $4.085$ to $4.600~\GeV$,''
JHEP \textbf{05}, 155 (2022) [arXiv:2112.06477 [hep-ex]].

\bibitem{BESIII_DsDs_2024}
M. Ablikim \textit{et al.} [BESIII Collaboration],
``Precise measurement of the $e^+e^-\to D_s^+D_s^-$ cross section at center-of-mass energies from threshold to $4.95~\GeV$,''
Phys. Rev. Lett. \textbf{133}, 261902 (2024) [arXiv:2403.14998 [hep-ex]].

\bibitem{Peng2024}
T.-C. Peng, Z.-Y. Bai, J.-Z. Wang, and X. Liu,
``Reevaluating the $\psi(4160)$ resonance parameter using $B^+\to K^+\mu^+\mu^-$ data in the context of unquenched charmonium spectroscopy,''
Phys. Rev. D \textbf{111}, 054023 (2025) [arXiv:2412.11096 [hep-ph]].

\bibitem{Salpeter1952}
E.~E. Salpeter and H.~A. Bethe,
``A relativistic equation for bound-state problems,''
Phys. Rev. \textbf{84}, 1232 (1951).

\bibitem{SalpeterOrigins}
E.~E. Salpeter,
``Bethe-Salpeter Equation -- The Origins,''
arXiv:0811.1050 [physics.hist-ph].

\bibitem{Chang2005}
C.-H. Chang, J.-K. Chen, and G.-L. Wang,
``Instantaneous formulation for transitions between two instantaneous bound states and its gauge invariance,''
Commun. Theor. Phys. \textbf{46}, 467--480 (2006) [arXiv:hep-th/0312250].

\bibitem{Chang2004}
C.-H. Chang, J.-K. Chen, X.-Q. Li, and G.-L. Wang,
``Instantaneous Bethe-Salpeter equation and its exact solution,''
Commun. Theor. Phys. \textbf{43}, 113--118 (2005) [arXiv:hep-ph/0406050].

\bibitem{Wang2006}
G.-L. Wang,
``Decay constants of heavy vector mesons in relativistic Bethe-Salpeter method,''
Phys. Lett. B \textbf{633}, 492--496 (2006) [arXiv:math-ph/0512009].

\bibitem{Micu}
L. Micu,
``Decay rates of meson resonances in a quark model,''
Nucl. Phys. B \textbf{10}, 521--526 (1969).

\bibitem{LeYaouanc}
A. Le Yaouanc, L. Oliver, O. P\`ene, and J.-C. Raynal,
``Hadron transitions in the quark model,'' (Gordon and Breach, Amsterdam, 1988).

\bibitem{Segovia2012}
J. Segovia, D. R. Entem, and F. Fernandez,
``Scaling of the $^3P_0$ strength in heavy meson strong decays,''
Phys. Rev. D \textbf{86}, 034026 (2012) [arXiv:1205.2215].

\bibitem{Wang2013Strong}
T. Wang, G.-L. Wang, H.-F. Fu, and W.-L. Ju,
``Two-body strong decay of $Z(3930)$ as the $\chi_{c2}(2P)$ state,''
JHEP \textbf{07}, 120 (2013) [arXiv:1305.1067 [hep-ph]].

\bibitem{Wan2026}
B.~D. Wan and S.~Q. Zhang,
``Nodal mechanism for the suppressed $D\bar D$ decay of $\psi(4040)$ in the Bethe-Salpeter framework,''
JHEP \textbf{07}, 254 (2026)
[arXiv:2605.10882 [hep-ph]].

\bibitem{WangGL2020}
G.-L. Wang, T.-F. Feng, and X.-G. Wu,
``Average speed and its powers $v^n$ of a heavy quark in quarkonia,''
Phys. Rev. D \textbf{101}, 116011 (2020) [arXiv:2003.10116].

\bibitem{Geng2019}
Z.-K. Geng, T.-H. Wang, Y. Jiang, G. Li, X.-Z. Tan, and G.-L. Wang,
``Relativistic effects in the semileptonic $B_c$ decays to charmonium with the Bethe-Salpeter method,''
Phys. Rev. D \textbf{99}, 013006 (2019) [arXiv:1809.02968].

\bibitem{Chang2015}
C.-H. Chang, H.-F. Fu, G.-L. Wang, and J.-M. Zhang,
``Some of semileptonic and nonleptonic decays of $B_c$ meson in a Bethe-Salpeter relativistic quark model,''
Sci. China Phys. Mech. Astron. \textbf{58}, 071001 (2015) [arXiv:1411.3428].

\bibitem{DingWan}
R. Ding, B.-D. Wan, Z.-Q. Chen, G.-L. Wang, and C.-F. Qiao,
``Finding $B_c(3S)$ states via their strong decays,''
Phys. Lett. B \textbf{816}, 136277 (2021) [arXiv:2101.01958 [hep-ph]].

\bibitem{Gui2018}
L.-C. Gui, L.-S. Lu, Q.-F. L\"u, X.-H. Zhong, and Q. Zhao,
``Strong decays of higher charmonium states into open-charm meson pairs,''
Phys. Rev. D \textbf{98}, 016010 (2018) [arXiv:1801.08791].

\bibitem{Belle_DsDsstar_2011}
G. Pakhlova \textit{et al.} [Belle Collaboration],
``Measurement of $e^+e^-\to D_s^{(*)+}D_s^{(*)-}$ cross sections near threshold using initial-state radiation,''
Phys. Rev. D \textbf{83}, 011101 (2011) [arXiv:1011.4397 [hep-ex]].

\end{thebibliography}
\end{document}